\documentclass[a4paper,10pt]{edpsci} 
\usepackage{amsmath}
\usepackage{graphicx}
\usepackage{epstopdf}
\usepackage[numbers]{natbib}
\usepackage{hyperref}
\usepackage{url}
\usepackage{subcaption}
\usepackage{booktabs}  
\usepackage{float}     
\usepackage{tikz}
\usetikzlibrary{matrix, positioning, calc, arrows.meta}

\usepackage{csquotes}

\makeatletter
\renewcommand{\@biblabel}[1]{[#1]}
\makeatother

\hypersetup{colorlinks=true,citecolor=blue,urlcolor=blue,linkcolor=blue}

\usepackage[ruled]{algorithm2e}
\usepackage[utf8]{inputenc}
\usepackage[english]{babel}
\usepackage{hyperref}
\usepackage{authblk}
\usepackage[normalem]{ulem}
\usepackage{booktabs}

\journalname{Journal Title}

\articletype{Research article}

\usepackage{amsmath}  
\usepackage{amssymb}  

\onecolumn

\title{ADIOSS — Automatic Diagnostic Of System Simulations}

\titlerunning{ADIOSS — Automatic Diagnostic Of System Simulations}

\author{Di Jiang\inst{1,2} 
\and
Sebastian Rodriguez\inst{1}
\and
Hervé Colin\inst{2}
\and
Yves Tourbier\inst{1,2}
\and
Francisco Chinesta\inst{1}
}

\authorrunning{D. Jiang et al.}

\institute{PIMM, Arts et Metiers Institute of Technology, Paris, France\and
Renault Group, Guyancourt, France}

\abstract{Automotive engineering makes extensive use of numerical simulation throughout the design process. The development of numerical models, their validation against experimental tests, and their updating during vehicle and engine projects constitute a core engineering activity. However, this activity must continuously evolve to reduce costs and lead times. 

In this context, we propose a method for detecting faulty modules within a system-level simulation workflow, represented as a graph of 0D models, following model updates. The proposed method requires a very limited number of system simulations and can therefore be easily integrated into existing engineering processes. It is designed as a toolbox based on well-established and widely validated techniques, including \textit{Dynamic Mode Decomposition}—commonly used for $3D$ model reduction, linear programming, and autoencoders.}

\keywords{0D vehicle simulation, Modular architecture, DMDc, Data-driven system identification, Module interaction analysis, Automotive simulation platform}

\newcommand{\etc}{\textit{etc}}
\newcommand{\DMD}{\textit{DMD}}
\newcommand{\DMDdetect}{\textit{MixED-DMD}}

\newcommand{\DMDc}{\textit{DMDc}}

\newcommand{\CTD}{\textit{CTD}}

\newcommand{\autoencoder}{autoencoder}
\newcommand{\NODYN}{\textit{NoDyn}}
\newcommand{\REM}[2]{\textbf{#1} : \textit{#2}}
\begin{document}

\maketitle

\tableofcontents

\newpage


\section{Introduction} 
Numerical simulation plays a central role in the automotive design process. The models used for decision-making are generally of two types: either $3D$ models based on the Finite Element Method\cite{zienkiewicz2005finite}, used when a design or computer-aided design ($CAD$) model is available, or $0D$ models, which rely on a controlled simplification of physics \cite{bordet2011modelisation0D}, \cite{hammadi2020reduction0D}, \cite{janiaud2011modelisation0D}. This article focuses specifically on $0D$ models, also referred to as \enquote{system models}, which we will call \enquote{modules} throughout this paper to avoid confusion with statistical models.

$0D$ models, or modules, are widely used to rapidly evaluate vehicle performance over various customer driving cycles. These modules dynamically simulate the evolution of a state $x$ according to a function $x_{t+1} = f(x_t)$ within complex computational chains, referred to as a workflow when several modules are connected through their input/output variables. This principle is illustrated by the simulation platform Siemens used by Renault \cite{siemens2023renault}.

The workflow is supplied by different engineering domains, each providing an updated and individually validated module. For instance, a workflow simulating the behavior of an electric vehicle over a given route may include modules for the battery, motor, inverter, battery cooling system, air conditioning system, and any component that consumes onboard energy when predicting energy consumption and driving range. A standard workflow includes more than ten modules. The module graph is executed by a system simulation platform that integrates them using so-called weak coupling\cite{blochwitz2011fmi}.

Our research focuses specifically on the validation process of system modules. In an automotive industrial context where validation timelines are very short and simulation budgets limited, it is essential that modules be reliable from their design stage. Experimental validation, often carried out at the end of a project, therefore requires a high level of quality from the modeling phase onward. The context is as follows:
\begin{itemize}
    \item A reference simulation workflow, denoted $W_0$, validated on customer driving cycles, is available. This workflow and its $m$ modules $M_0^1...M_0^m$ were, for example, validated at the end of the previous vehicle project; the results were compared with final vehicle validation tests and were deemed satisfactory;
    
    \item A new version $W_1(M_1^1...M_1^m)$ is delivered by the various engineering teams and external suppliers; the solvers or even the platform itself may also have been updated. This new workflow does not produce the same results as the previous one on customer cycles, for example with respect to a global criterion such as energy consumption, driving range, thermal comfort, \textit{etc.} Since the previous results serve as the reference, discrepancies must be explained before using the new version in a future vehicle project. A module contributing to these discrepancies is referred to as \enquote{faulty} in the remainder of this article.
\end{itemize}

Many changes may occur between two versions of a workflow, and their impact on performance criteria is not always obvious. Moreover, external suppliers do not detail all implemented changes and do not provide access to their source code. Because the workflow is used within ongoing projects, high agility is required in the validation and error-correction process. It must be possible to quickly detect the source of a discrepancy in order to promptly inform the relevant engineering team or supplier and ensure that project teams always use a validated version. This issue is further complicated by the fact that end users of the modules are not necessarily experts in numerical simulation. A module developed by a specialist is often reused in other contexts, sometimes without a deep understanding of its limitations. This situation highlights the need for an automated tool capable of efficiently identifying faulty modules within a complex simulation workflow, \textbf{at minimal cost and within short timeframes}.

\begin{figure}[H]
    \centering
    \includegraphics[width=0.35\textwidth]{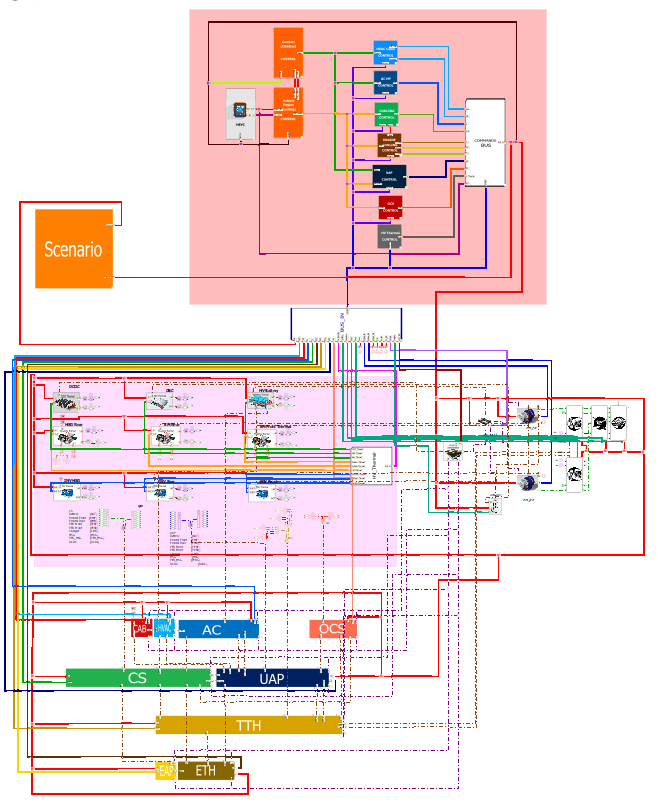}
    \caption{An example of a simulation workflow. Each box represents a module, implemented in \textit{Amesim}, \textit{Simulink}, \textit{Matlab}, or as a map-based model. The module graph is represented here using the platform formalism. The lines represent input/output variables, energy flows, control signals, \textit{etc.}}
    \label{fig:illustration_workflow}
\end{figure}

\subsection{Notations used}

A system simulation consists of executing a graph of functions at each time step. The functions, \textbf{called modules}, take as inputs variables of very different natures:

\begin{itemize}
    \item Physical variables whose values are governed by equations. In this article, these variables must be modeled using a statistical method;
    \item Thresholds applied to physical quantities computed within the workflow and used for control purposes, the control itself being implemented as a module. For example, adaptive cruise control activates above $30\,km/h$;
    \item Control variables or variables defining the customer driving cycle, whose values are imposed over time and therefore do not need to be modeled. For example, road slope, outside temperature, speed setpoint of the cycle, \etc{}.
\end{itemize}

The remainder of the article uses the following notations:

\begin{itemize}
    \item The time step is denoted $t$ and varies from $1$ to $n$;
    \item The reference modules are denoted $M_0^j$, with $j$ ranging from $1$ to $m$;
    \item The module $M_1^j$ is the new version of $M_0^j$. The objective is to determine whether it is faulty, that is, genuinely different from $M_0^j$;
    \item The variables of module $M^j$ are denoted $x_{j,i}$ (or $x_{t,j,i}$ when the time index is explicitly considered). When the module structure is not distinguished, the variables of the entire workflow are simply denoted $x_i$, with $i$ ranging from $1$ to $p$, including all physical variables, control variables, and driving cycle definition variables;
    \item The set of variables internal to the modules are represented by the state vector $\overline{x_t}$ at time step $t$. During the simulation, at each time step, the value of each variable $x_{t,i}$ (or $x_{t,j,i}$) is collected at the input or output of each module;
    \item A simulation produces a data table denoted $T_{t,i}$, where time is arranged in rows (often in columns in the publications cited in our bibliography);
    \item $T_0$ denotes the data table resulting from the reference simulation $W_0$;
    \item ``DoE'' denotes a design of experiments on Boolean variables. $\overline{DoE}$ denotes the same design in which all $0$ and $1$ values are swapped.
\end{itemize}

Due to the complexity of the simulation platform and the need for accurate results, computation time is significantly longer than real time. For example, reproducing a customer driving cycle lasting 30 minutes, with a sampling time of 0.5 seconds, can require more than ten hours. The workflow validation process involves numerous cycles representing very different situations or regulatory requirements. The proposed method must therefore be economical in terms of the number of simulations in order to enable systematic use.

\section{Modular Validation and Sensitivity Analysis} 
\subsection{Validating One Module at a Time}
The first method considered, which naturally comes to mind, is to validate the modules one by one. 
This simply requires running simulations in which only one module differs from the reference configuration. 
Let $T_0$ denote the trajectory of system outputs obtained from the reference simulation, and $T_j$ the trajectory obtained when module $M^j$ is replaced by its updated version while all other modules remain identical to the reference. 

This approach requires $m+1$ simulations: the reference simulation plus one simulation per module. 
The method is easy to implement, and simulations can be run in parallel to reduce turnaround time. 
It allows detection of even very small variations resulting from the replacement of module $M^j$ by computing the difference $T_0 - T_j$. 
This makes it possible to determine whether a change has actually been introduced.

However, modules interact strongly. Differences are always observed as soon as a module is modified, but it is not possible to rank module changes in decreasing order of importance. Moreover, detection capability depends on the simulated driving cycle. A modified threshold that is not reached during the simulation will not be detected (which is of course true for any method). The opposite situation can also occur: a minor modification may have a significant effect because it is not compensated by the update of another module. Most of the time, modules are updated but the resulting discrepancy is small; what truly matters is their interaction with other modules. This first method detects large variations and highly sensitive modules. It should be noted that system simulation workflows include only sensitive modules; modules without impact are simply not integrated into the workflow.

\subsection{Sensitivity Analysis}

A first algorithm was developed to detect the sources of discrepancies, based on the design of experiments methodology \cite{livreyto}, as are most algorithms relying on computationally expensive simulations.

The detection of faulty modules can be formulated as a combinatorial search over Boolean variables. Let $X_j$ be a Boolean variable equal to $1$ when module $M_0^j$ is replaced by $M_1^j$ in $W_0$, and $0$ otherwise. Similarly, let $Z_j$ be a Boolean variable equal to $1$ if $M_1^j$ is replaced by $M_0^j$ in $W_1$. We assume that modules are the sole cause of the discrepancy $W_1 - W_0$. The objective is to find the vectors $X$ and $Z$ that best explain $W_1 - W_0$. The search may be performed using $X$ alone, $Z$ alone, or both. If $X = Z$ but the two simulation results differ, then the modules do not fully explain the discrepancy $W_1 - W_0$. In that case, purely numerical causes must be investigated, such as replacing the platform itself (which could also be represented by a Boolean variable), or changing the solver used for a given module.

The simplest algorithm uses only $X$ and corresponds to applying design of experiments to search for an optimum:

\begin{enumerate}
    \item Assign a Boolean variable to each module;
    \item Choose a $DoE$.
    \item Run the reference simulation $W_0$;
    \item Run the $N$ simulations defined by $DoE$;
    \item Compute the global performance criterion to be explained for each experiment in the design, for example the predicted driving range of the electric vehicle. This criterion is denoted $W_0(X_i)$;
    \item Construct the response vector, $\forall \,t \, \in\, 1..N,\quad Y_t = W_0(X_t) - W_0$;
    \item Build the regression model $\hat{Y_t} = f(X_t)$ that explains the deviation from the reference as a function of module replacements. Model construction includes variable selection;
    \item Use the regression model to determine the smallest $X_{opt}$ that minimizes the discrepancy $W_1 - W_0$. The smallest $X_{opt}$ minimizes the number of Boolean variables equal to $1$;
    \item Validate the solution by running simulation $X_{opt}$ and computing $Y_{opt}$;
    \item If $Y_{opt}$ differs from $\hat{Y}_{opt}$, then some terms are missing in the statistical model. Additional experiments must be added, return to step $4$, and select a more complex statistical model (including additional interaction terms).
\end{enumerate}

The search for the best regression model and for $X_{opt}$ can be merged by taking into account the specific structure of the problem. Since the variables $X_j$ are Boolean, that is qualitative, interpolation has no meaning. The set of possible regression models is of the form

$$
Y = \alpha_0 + \sum_j \alpha_j.X_j + \sum_{j_1,j_2}\alpha_{j_1,j_2}.X_{j_1}.X_{j_2}+...\alpha_{j_1..j_m}.\prod_{j=1}^P X_j
$$

The maximum number of non-zero coefficients $\alpha_{j_1...}$ is $2^m$ if the workflow contains $m$ modules.  
\textbf{The model is necessarily a multiple linear regression model}. There is no need to use any other regression method.

For example, if only modules $M_1$ and $M_2$ are responsible for the discrepancy $W_1 - W_0$, then the regression model reduces to

$$
Y = \alpha_0 + \alpha_1.X_1 + \alpha_2.X_2 + \alpha_{1,2}.X_1.X_2
$$

\textit{Remark: If the deviation from the reference is modeled and a full factorial design is used, then $\alpha_0$ is equal to zero, since the reference configuration is itself one of the experiments in the full factorial design.}

The case where $\alpha_{1,2}$ is zero while both modules are required to explain the discrepancy is highly unlikely. It would correspond to two modules that do not share any variables in the workflow during the simulation, equivalent to two independent workflows running in parallel. This situation does not occur in practice; the interaction term $X_1.X_2$ is necessary. This property has a dramatic consequence: for $P$ modules to be detected among $m$, the regression model to be estimated includes $2^P$ coefficients among the $2^m$ possible ones if the discrepancy $W_1 - W_0$ is to be perfectly explained, which implies a cost of $2^m$ simulations. In this context, \textbf{only the full factorial design guarantees detection of all faulty modules}, leading to a theoretically unacceptable computational cost.

In practice, faulty modules are usually known or suspected, and the entire workflow is not modified at once. Moreover, the initial objective is detection rather than detailed explanation. If the number $P$ of faulty modules is fixed without knowing which ones, then the experimental design must contain all full factorial designs of $P$ variables among $m$, which remains very costly.

The problem of searching for the best regression model shares similarities with the \CTD{} (Combinatorial Test Design, \cite{mandl1985use}) methodology and with the search for failure scenarios in autonomous vehicles and advanced driver assistance systems (ADAS). Numerous methods have been developed \cite{kuhn2013introduction}, and their presentation is beyond the scope of this article. The first studies date back to the 1990s and research is still ongoing. Their application has resulted in high simulation costs and motivated our work.

However, our problem differs from classical software testing. In \CTD{}, experiments are independent, and the objective is to identify those that trigger an error. In our case, experiments are dynamic simulations, where inputs and outputs interact over time. Our problem is closer to software validation when validating a sequence of actions (see the control variables discussed later in the article).

\section{Case Study} 
\label{CaseStudy}
\subsection{Workflow Description}

The test case selected to evaluate the proposed methodology is based on an open-source model developed by MathWorks \cite{mathworks2022bev}, representing the propulsion system of an electric vehicle.

\begin{figure}[H]
    \centering
    \includegraphics[scale=0.2]{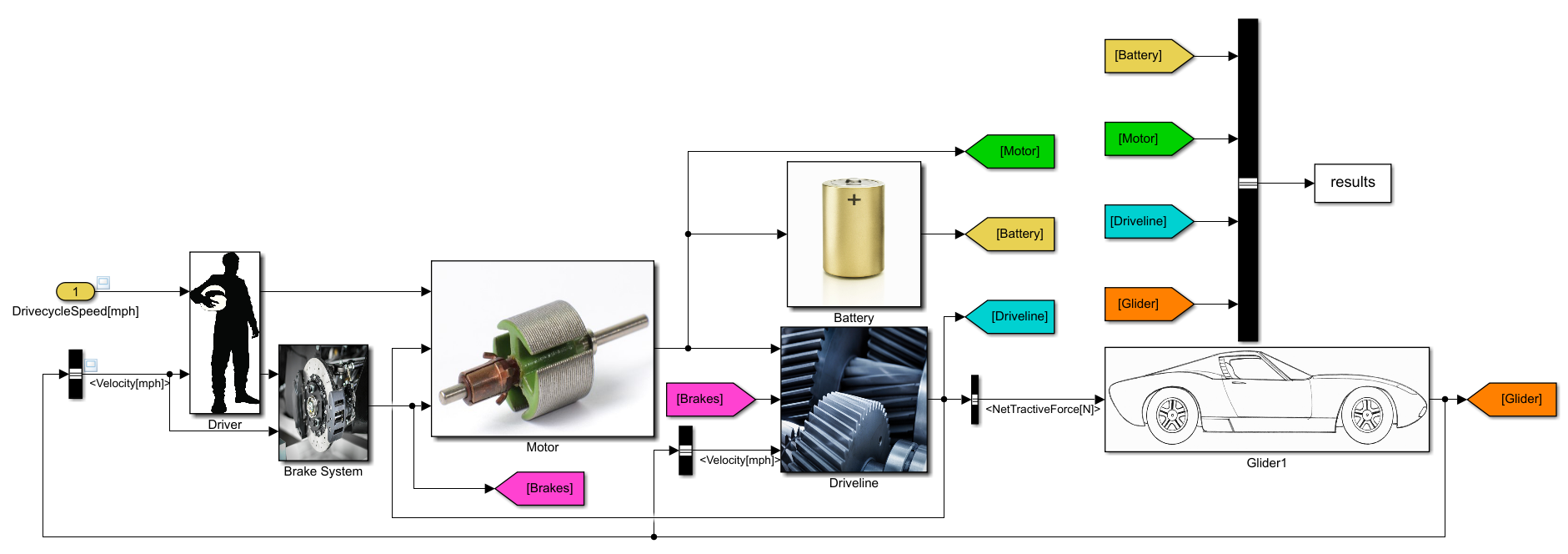}
    \caption{Simulink model of an electric vehicle powertrain.}
    \label{fig:exemple_model_simulink}
\end{figure}

The workflow (see Figure~\ref{fig:exemple_model_simulink}) is organized in a modular manner and driven by an imposed driving cycle. This cycle defines the temporal evolution of the vehicle’s target speed (see Figure~\ref{fig:vitesse_cible_variable_impose}).

\subsection{Representation of a Module Version Change}

In this study, a module version change is modeled through a controlled variation of a representative internal parameter. This approach makes it possible to emulate the effect of a local model modification while preserving a strictly identical overall architecture, which is a necessary condition for consistent comparison of simulations. For each module under study, a key parameter is selected due to its direct influence on longitudinal dynamics or vehicle energy consumption.

\begin{figure}[ht]
\centering
\begin{subfigure}[t]{0.48\textwidth}
    \centering
    \includegraphics[width=\textwidth]{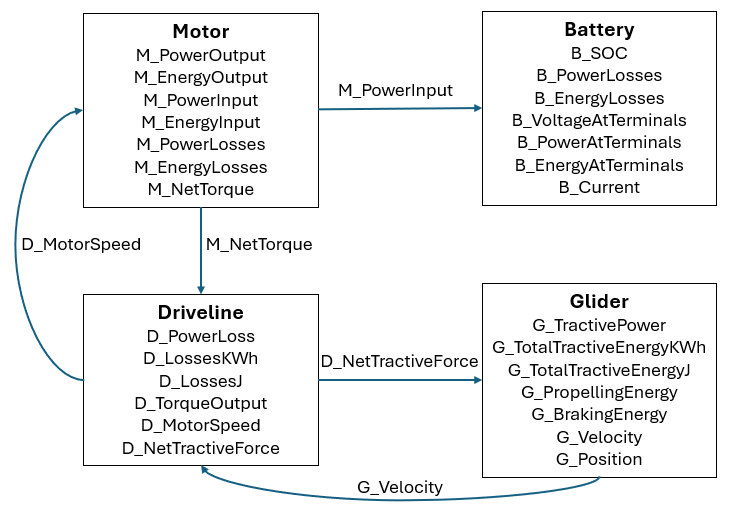}
    \caption{Simplified call graph of modules and variable flows.}
    \label{fig:graphe_appel_module_new}
\end{subfigure}
\hfill
\begin{subfigure}[t]{0.48\textwidth}
    \centering
    \includegraphics[width=\textwidth]{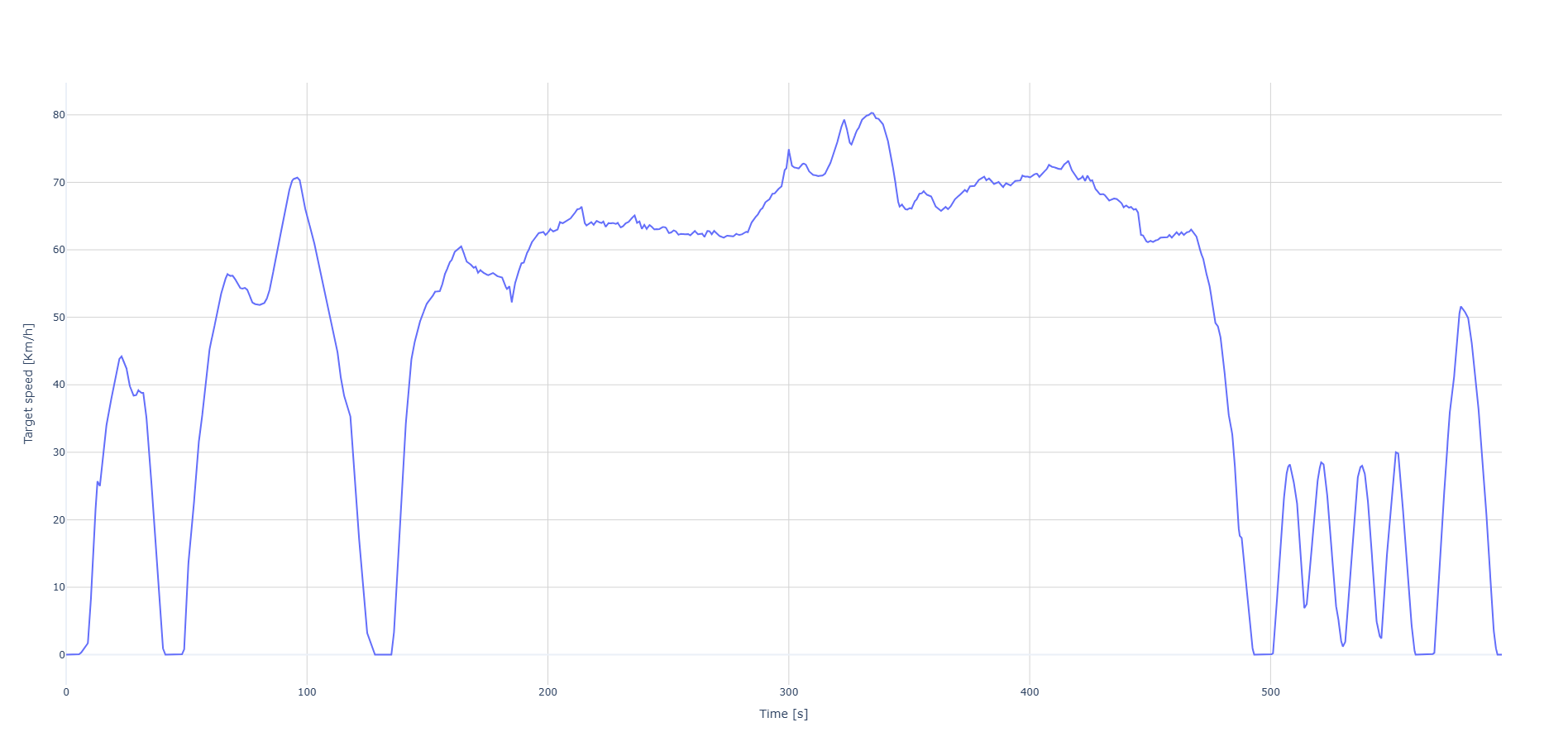}
    \caption{Evolution of the imposed vehicle target speed.}
    \label{fig:vitesse_cible_variable_impose}
\end{subfigure}
\caption{Workflow organization and imposed variable of the test case.}
\end{figure}

Figure~\ref{fig:graphe_appel_module_new} presents a simplified call graph of the workflow, in which only interactions between modules whose versions may vary are represented. This graph is a simplified version of Figure~\ref{fig:exemple_model_simulink} and highlights the flows exchanged between modules.

\subsection{Level of Variation of Key Parameters}

An exploratory study was conducted to define relevant perturbation levels. Uniform variations ranging from $5\%$ to $30\%$ were applied to the selected parameters. The impact of these variations was evaluated using the sum of squared differences between the outputs of the modified workflow and those of the reference one.

Two configurations were retained for all simulations presented in this article. The \emph{Battery} module is modified by increasing its internal resistance by $10\%$, while the \emph{Motor} module is modified by increasing its maximum output torque by $10\%$. 
This choice makes it possible to construct a test case including both a module whose impact is easily detectable (\emph{Motor}) and a module that is more difficult to identify (\emph{Battery}). 
To justify this selection, a preliminary sensitivity study was conducted in which different percentages of parameter modifications were applied to each module. 
The resulting simulations were analysed by comparing the trajectories of key variables with those of the reference simulation using Euclidean distance metrics, as well as domain-specific performance indicators such as the vehicle autonomy. 
This analysis showed that modifications in the Motor module produce larger and more easily detectable deviations, whereas changes in the Battery module tend to induce more subtle variations in the system behaviour.
The other modules, \emph{Driveline} and \emph{Glider}, should not be detected since they have not been modified.

\subsection{Imposed Variable}

The driving cycle constitutes the main imposed variable of the test case. It defines the temporal evolution of the vehicle’s target speed and is identical for all simulations, enabling comparison of the dynamic responses of the different configurations.

\begin{table}[H]
    \centering
    \scriptsize
    \renewcommand{\arraystretch}{1.2}
    \begin{tabular}{|p{3cm}|p{4cm}|p{7cm}|}
        \hline
        \textbf{Module} & \textbf{Parameter} & \textbf{Description} \\
        \hline
        Motor & Maximum output torque & Maximum torque delivered by the motor, influencing acceleration capability and energy recovery. \\
        \hline
        Battery & Internal resistance & Internal battery resistance, determining electrical losses and overall energy efficiency. \\
        \hline
        Driveline & Gear ratio & Ratio governing the trade-off between torque and speed, influencing mechanical performance. \\
        \hline
        Glider & Rolling resistance coefficient & Parameter affecting energy consumption at constant speed. \\
        \hline
    \end{tabular}
    \caption{Studied modules and selected representative parameters}
    \label{tab:modules_parametres}
\end{table}

\section{Single-Simulation Method with Reference (\DMDdetect{})}
The proposed method is derived from \DMD{} (Dynamic Mode Decomposition)\cite{kutz2016dynamic,Schmid2010} and combines multiple dynamic matrices.  
The objective is to select the subset of modules that best explains the data. It represents the minimal cost in terms of number of simulations: the reference simulation plus a single additional simulation in which the module versions are modified according to $DoE$. The method can also be applied to multiple simulations to increase detection capability.

Sensitivity analysis is \enquote{classical} in the sense that it relies on well-established tools, but its simulation cost quickly becomes prohibitive for project teams operating under tight deadlines. \textbf{To reduce this cost, we implement a very simple idea: changing the module versions, the $X_j$ and $Y_j$, during the simulation itself}. The $X_j$ and $Y_j$ variables are similar to control variables and become $X_{t,j}$ and $Y_{t,j}$. The proposed method is described using only the $X_{t,j}$ variables but remains applicable with the $Y_{t,j}$ variables. This simulation is denoted $W_0(DoE)$ and produces the data table $T_0(DoE)$.

The objective is to take advantage of the large number of time steps in customer driving cycles (from $4000$ to $18000$) to maximize the size of the experimental design.

A $DoE$ and its implementation strategy must be defined. For example:

\begin{itemize}
    \item A full factorial design on $7$ modules with $128$ experiments yields $128$ replacement sequences, each lasting $100$ consecutive time steps;
    \item The same $128$-experiment design repeated twice with sequences of $50$ time steps;
    \item The full factorial design of $2^{10} = 1024$ experiments for $10$ modules, implemented in sequences of $11$ time steps;
\end{itemize}

The number of time steps in a sequence must be sufficient to reveal the effects of module replacement. The experiments in the design may be performed in any order, even randomly. The execution order changes the result $W_1 - W_0(DoE)$ due to dynamic effects. If the workflow dynamics are complex, or if high-order interactions between modules must be detected, then the full factorial design should ideally be used.

In the presented example, the total simulated duration is $596\sec{}$, with a time step of $0.1\sec$; the resulting data table therefore contains $5960$ time steps. Simulations are performed by modifying module versions according to a four-factor full factorial design, corresponding to $16$ combinations of module versions.

Each experiment is simulated over $20$ time steps, and the experimental design is repeated successively $18$ times in order to cover the total simulation duration (plus part of the $19^{\text{th}}$ repetition).

A second experimental design was also tested: the same $16$ combinations were implemented twice, with $186$ time steps per combination. The results obtained with this second design lead to the same conclusions as those of the first and are therefore not detailed here.

As in sensitivity analysis, the objective is to explain the data, \textbf{but this time in a detailed manner rather than aggregating it into a single scalar}. Up to four simulations may be available for each customer driving cycle: $W_0$, $W_1$, $W_0(DoE)$ and $W_1(\overline{DoE})$.  
Each of these simulations produces a data table $T$ with $18000$ rows (time steps) and $60$ columns (variables). For simulations $W_0(DoE)$ and $W_1(\overline{DoE})$, the table additionally contains the columns of $X_{t,j}$ controlling module versions, since they are imposed by the experimental design.

The workflow is a dynamic model, meaning that $x_{t+1} = f(x_t)$. We adopt the \DMD{} formalism by assuming linear dynamics:

\begin{equation}
x_{t+1} = A.x_t
\end{equation}

All subsets of size $q$ modules (or up to size $q$ for a more general problem) are defined and denoted $\mathcal{M}_k$. For each $\mathcal{M}_k$, we associate the $2^q$ matrices $A_{k,l}$ corresponding to all terms of the complete linear model over the $q$ modules in $\mathcal{M}_k$. For the subset \{$Battery$, $Motor$\}, for example, this yields three matrices corresponding to $Battery$, $Motor$, and $Battery \times Motor$.

\begin{equation}
\begin{aligned}
& \underset{A, A_{k,l}, \epsilon}{\text{Min}}  \quad P_{\epsilon}. \sum_{t=1}^{n} \|\epsilon_t\|_2^2 + P_a .\|A\|_1 + \sum_{k,l}P_{k,l} .\|A_{k,l}\|_1 \\
& \text{under  :} \quad x_{t+1} =  A.x_t + \sum_{k,l \in \mathcal{T}_{all}} \delta_{t,k,l} \times A_{k,l}. x_t + \epsilon_t \\
\end{aligned}
\end{equation}

Where:
\begin{itemize}

    \item $A_{k,l}$ denotes a corrective transition matrix associated with the subset of modules $\mathcal{M}_k$. 
    The index $k$ identifies the considered combination of modules, while $l$ indexes the different terms associated with this combination after embedding.

    \item $\delta_{t,k,l} = \prod_{j \in \mathcal{M}_k} \text{DoE}_{t,j}$ is a known constant representing the activation of term $k,l$ at time $t$;

    \item The symbol $\times$ in $\delta_{t,k,l}\times  A_{k,l}.x_t$ denotes an element-wise product;

    \item $P_{\epsilon}$, $P_a$, $P_{k,l}$ are positive scalar constants representing the weights of the residuals, the reference matrix $A$, and the matrices $A_{k,l}$, respectively. These weights must be chosen to achieve lexicographic optimization: we seek the smallest matrices $A$ and $A_{k,l}$ corresponding to the minimal residual solution. It is required that $P_{k,l} > P_a$ to minimize corrections applied to $A$.\newline

    \item In particular, the interaction matrix associated with a pair of modules (for example $A_{\text{Battery,Motor}}$) must be interpreted as a higher-order correction term with respect to the individual correction matrices $A_{\text{Battery}}$ and $A_{\text{Motor}}$. Therefore, its penalization weight must be chosen larger than the penalization weights of the corresponding individual matrices (i.e., $P_{\text{Battery,Motor}} > P_{\text{Battery}}$ and $P_{\text{Battery,Motor}} > P_{\text{Motor}}$), in order to favor simpler explanations based on single-module effects before introducing interaction terms.

\end{itemize}

In this formulation, the matrices $A_{k,l}$ act as minimal corrections to matrix $A$ for time steps corresponding to updated module versions. 
The computation is performed combination by combination; in the presented test case described in Section~\ref{CaseStudy}, this requires one optimization per pair of modules. 
A more global approach, consisting of identifying the best combination in a single optimization, requires a more complex \textit{PL-MIP} formulation \cite{Nemhauser1988,Bertsimas1997} but is less efficient in practice. 
In the presented test case described in Section~\ref{dmdavecembedding}, finding the best pair of modules requires $12\,\mathrm{s}$ excluding the \autoencoder{} step, which computes the nonlinear embedding of the reference data into a latent space. 
The \autoencoder{} is trained on the reference simulation data and consists of an encoder that maps the high-dimensional system states to a lower-dimensional latent representation, and a decoder that reconstructs the original variables from this latent space. 
This latent representation is then used as input for the \DMDdetect{} procedure, allowing the method to capture nonlinear structures in the data while keeping a linear dynamical model in the reduced space. 
Our formulation solves the global problem using a simple loop that is easily parallelizable. 
Our Python implementation relies on the \textbf{cvxpy} package \cite{Diamond2016} for linear/quadratic programming (the problem can be solved using $L_1$ or $L_2$ norms).

\REM{Remark 1}{The data table consists of the reference table $T_0$, in which all module control variables are set to $0$, and the table resulting from the experimental design $T_0(DoE)$. This allows a very high weight to be assigned to the reference dynamics and reinforces the \enquote{corrective} nature of the module dynamics.}

\REM{Remark 2}{The \DMDdetect{} method can also be applied to physical experiments, provided that module versions can be modified during the test. This is possible for control modules, maps, or software sensors, for example. However, it is obviously not possible to change the motor during a physical test. The $L_1$ version is the \enquote{robust} version, particularly suitable for experiments as it is more resistant to measurement noise. The $L_2$ norm is more sensitive and better suited for detecting very slightly faulty modules, but this also implies a higher rate of false detections.}

\subsection{Reference Model Identification Using \DMDc{}}
Before applying \DMDdetect{}, we assess the detection capability of classical \DMD{}. This detection capability depends on the quality of the modeling. We evaluate the relevance of classical \DMD{} by applying it to the reference data table alone. In practice, \DMDc{} (introduced by Proctor et al.~\cite{Proctor2016}) must be used because the imposed variable (\textit{speed\_setpoint}) is a control variable. The dynamic model becomes:

\begin{equation}
x_{t+1} = A.x_t + B u_t
\end{equation}

where $u_t \in \mathbb{R}^{m+1}$ is the control vector representing the \textit{speed\_setpoint} and the module activations (Boolean variables), and $B \in \mathbb{R}^{(m+1) \times p}$ is the matrix describing their influence on the states.

\begin{figure}[H]
  \centering

  \includegraphics[
      width=0.95\textwidth,
      trim={0cm 35cm 0cm 4cm},
      clip
  ]{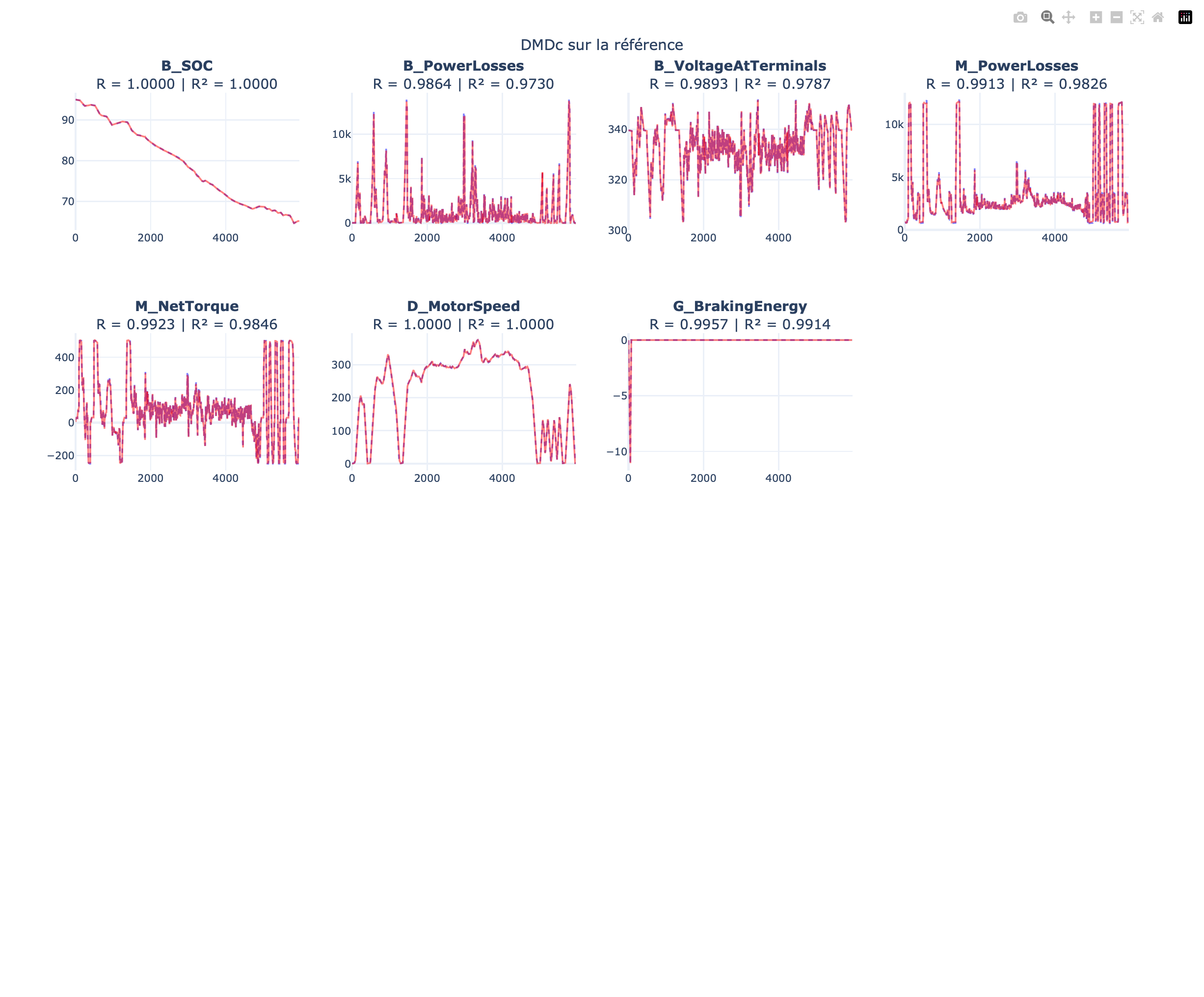}

  \vspace{0.8em}

  \includegraphics[
      width=0.95\textwidth,
      trim={0cm 35cm 0cm 4cm},
      clip
  ]{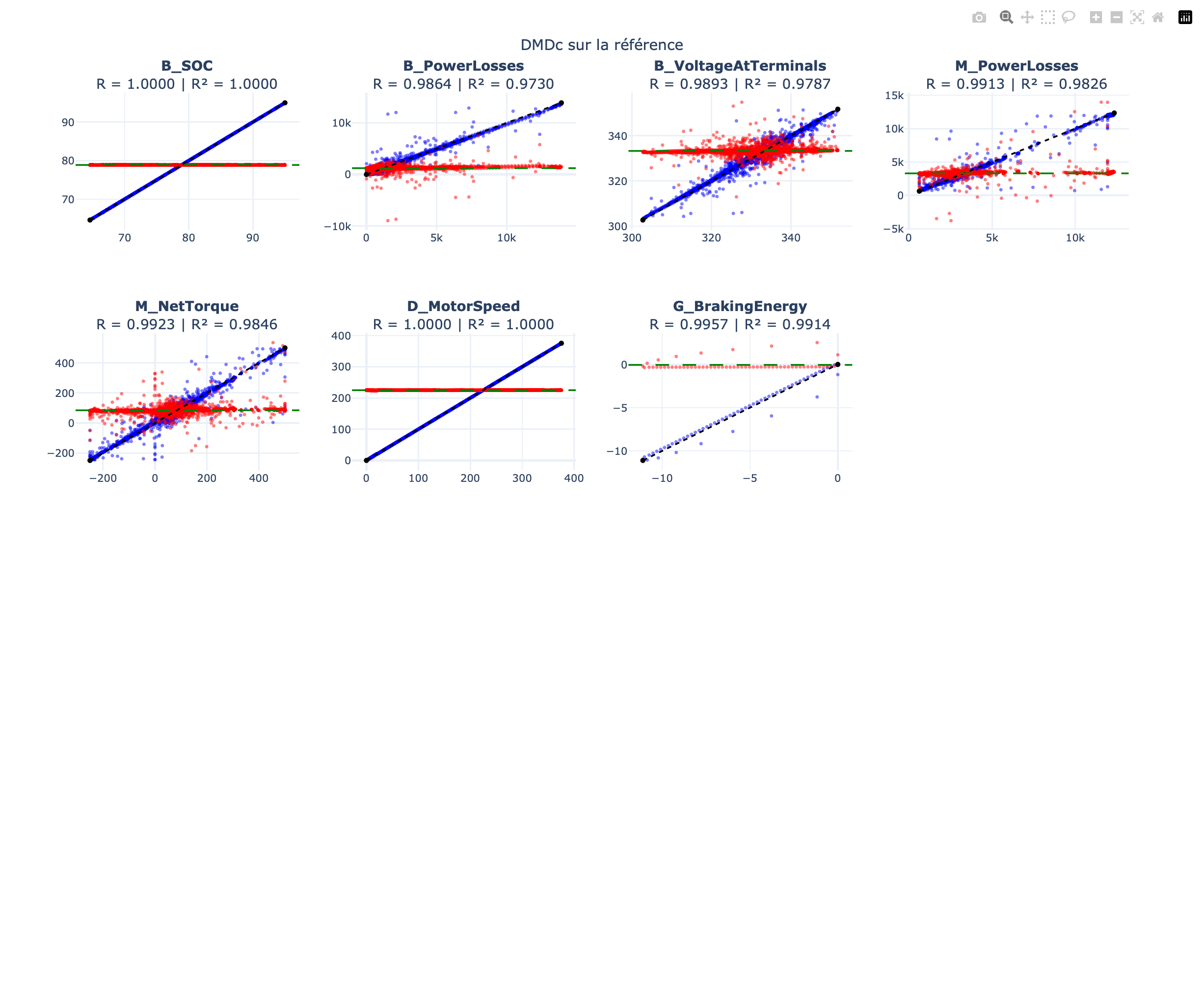}

  \caption{\DMDc{} applied to the reference data table. In the time-series plots, the models appear almost perfect. However, in the standard quality plots, it can be observed that $5$ out of $7$ models are imperfect. The blue series represents the predicted response as a function of the true response. The red series represents the residual plus the mean of the response. The underlying dynamics are almost linear, but not entirely. This justifies the use of embedding (see Section~\ref{dmdavecembedding}) and may also explain detection issues.}
  \label{graf:dmdcRefQualite}
\end{figure}

The \DMDc{} method using the $L_2$ norm can only be applied to the reference data after removing redundant variables; ultimately, only $7$ out of the initial $28$ variables remain. 
These $28$ variables correspond to the main signals exchanged between modules and are highlighted in Fig.~\ref{fig:graphe_appel_module_new}, which shows the simplified call graph of modules and variable flows. 
The results obtained using the remaining $7$ variables are detailed in Figures~\ref{fig:dmdc_results} and \ref{graf:dmdcRefQualite}. 
The dynamics appear almost linear, but this will prove insufficient to achieve good detection performance.

The spectral radius $\rho(A) = 1.0101$ is slightly greater than one, indicating that the identified linear model is marginally unstable. Although the deviation from unity is small, it may lead to progressive error growth during long simulations and partly explains the limitations of the classical \DMD{} approximation.

\begin{table}[H]
    \centering
    \vspace{1em}
    \textbf{Spectral radius} $\rho(A) = \max |\lambda_i| = 1.0101$
    \smallskip

    \begin{minipage}[t]{0.58\textwidth}
        \centering
        \textbf{Matrix A (Internal dynamics)}\\
        \resizebox{\textwidth}{!}{
        \begin{tabular}{lrrrrrrr}
            \hline
            & \rotatebox{90}{B\_SOC} & \rotatebox{90}{B\_PowerLosses} & \rotatebox{90}{B\_VoltageAtTerminals} & \rotatebox{90}{M\_PowerLosses} & \rotatebox{90}{M\_NetTorque} & \rotatebox{90}{D\_MotorSpeed} & \rotatebox{90}{G\_BrakingEnergy} \\
            \hline
            B\_SOC                &  1.000 &  0.000 &  0.001 & -0.000 &  0.000 &  0.000 &  0.000 \\
            B\_PowerLosses        &  0.001 &  0.965 & -0.008 &  0.004 & -0.007 & -0.566 &  0.000 \\
            B\_VoltageAtTerminals & -0.003 &  0.011 &  0.986 & -0.014 &  0.004 &  0.235 & -0.000 \\
            M\_PowerLosses        & -0.001 & -0.003 &  0.024 &  0.974 &  0.030 & -0.229 &  0.003 \\
            M\_NetTorque          &  0.002 & -0.012 & -0.004 &  0.001 &  0.991 & -0.134 &  0.002 \\
            D\_MotorSpeed         &  0.000 &  0.000 &  0.001 & -0.003 &  0.011 &  0.995 & -0.000 \\
            G\_BrakingEnergy      & -0.001 &  0.001 &  0.002 & -0.000 &  0.002 &  0.003 &  0.995 \\
            \hline
        \end{tabular}}
    \end{minipage}
    \hfill
    \begin{minipage}[t]{0.38\textwidth}
        \centering
        \textbf{Matrix B (Control)}\\
        \resizebox{\textwidth}{!}{
        \begin{tabular}{lr}
            \hline
                          & speed\_setpoint \\
            \hline
            B\_SOC                & -0.000 \\
            B\_PowerLosses        &  0.564 \\
            B\_VoltageAtTerminals & -0.236 \\
            M\_PowerLosses        &  0.225 \\
            M\_NetTorque          &  0.125 \\
            D\_MotorSpeed         &  0.005 \\
            G\_BrakingEnergy      & -0.002 \\
            \hline
        \end{tabular}}
    \end{minipage}

    \caption{\DMDc{} model of the reference workflow.}
    \label{fig:dmdc_results}
\end{table}

\subsection{Module Detection Using \DMDdetect{} Without Latent Embedding}
\begin{figure}[H]
  \centering
  \includegraphics[
      width=0.85\textwidth,
      trim={1cm 1.5cm 2.5cm 4cm},
      clip
  ]{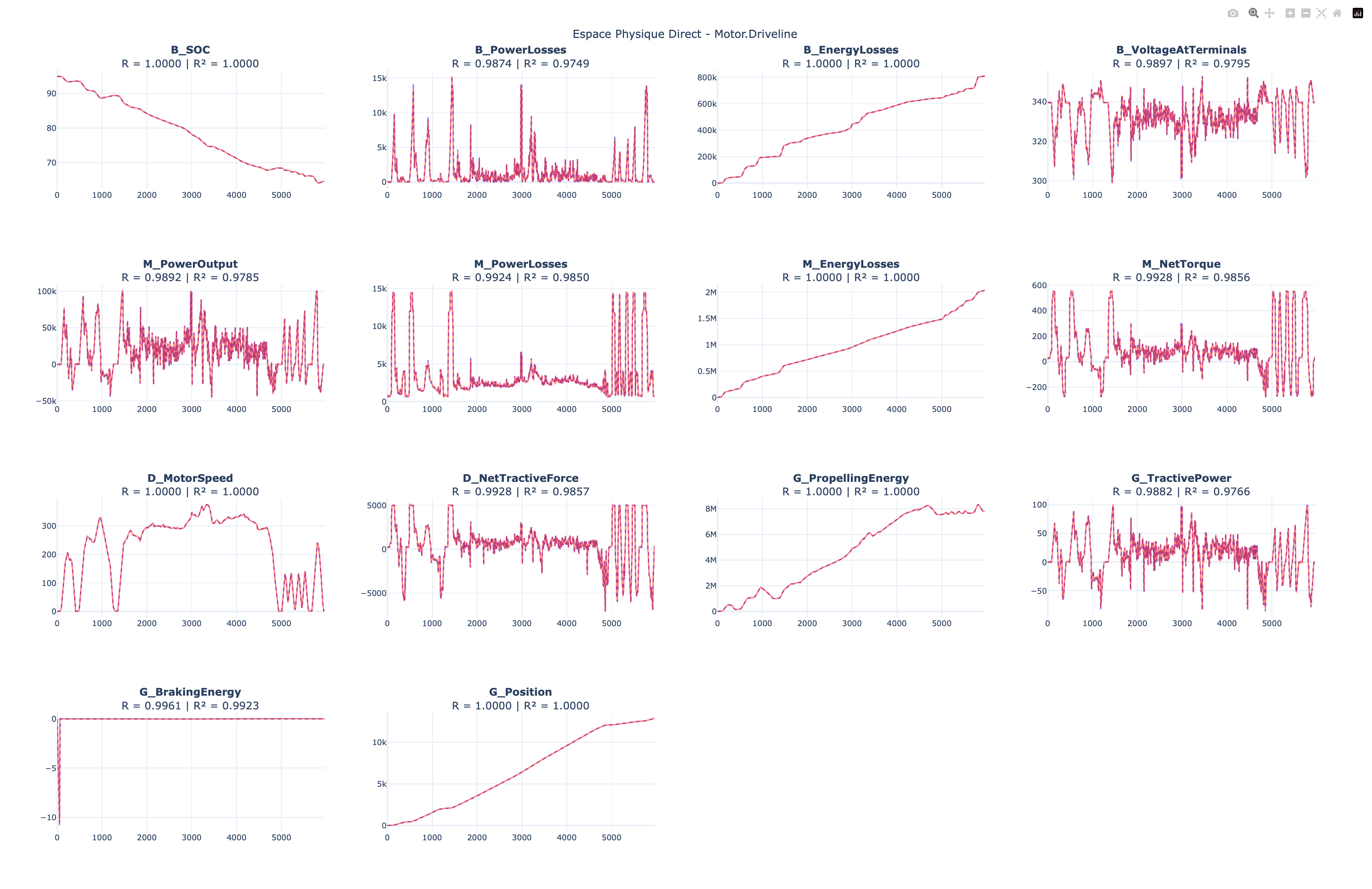}
  \caption{Selection of the best combination of two modules, without embedding. 
  The models are nearly perfect; the measured and predicted curves are almost everywhere superimposed.}
  \label{graf:DMDcombinaisonSelectionNoEmbedding}
\end{figure}

\begin{figure}[H]
  \centering
  \includegraphics[
      width=0.85\textwidth,
      trim={1cm 1.5cm 2.5cm 4cm},
      clip
  ]{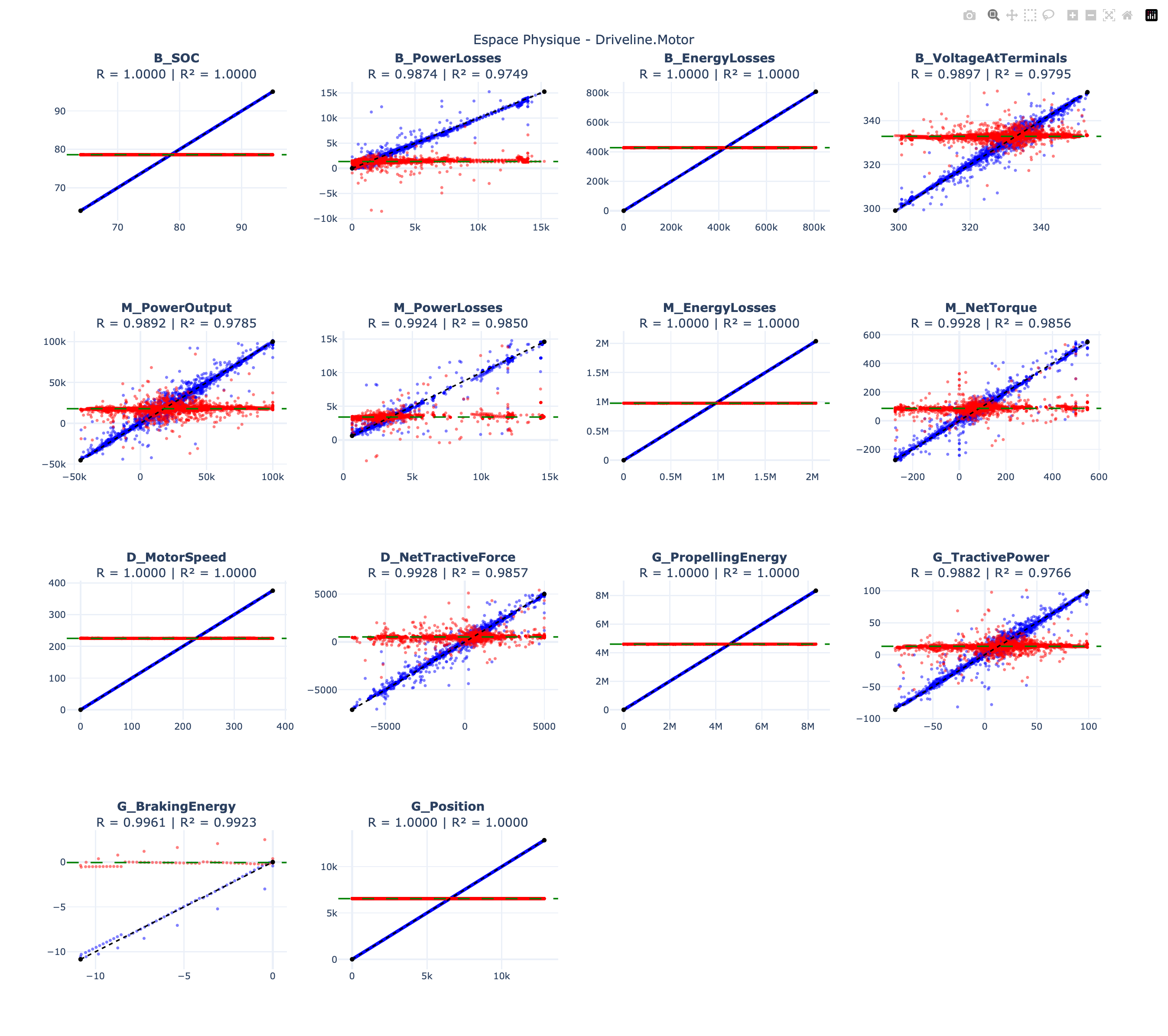}
  \caption{Selection of the best combination of two modules, without embedding. 
  The models appear nearly perfect in time-based predictions (see Figure~\ref{graf:DMDcombinaisonSelectionNoEmbedding}), 
  but are in fact far from perfect across all variables. The overall improvement compared to the reference \DMD{} model is minimal.}
  \label{graf:dmdSelectionNoEmbeddingYYp}
\end{figure}

 The first application, without embedding and using the $L_1$ norm, is performed on the $14$ remaining variables after greedy selection of the least correlated variables. The detailed results are shown in Figures~\ref{fig:DMD2MatrixAlpha} and \ref{graf:DMDcombinaisonSelectionNoEmbedding}. The \DMD{} method with the $L_1$ norm handles correlated variables reasonably well due to the $PL$ objective, which minimizes the absolute values of the elements of matrices $A$ and $A_{k,l}$; a variable that is strongly correlated with others tends to have its coefficients driven to zero. The method correctly detects the most influential version change, \emph{Motor}, but it makes a slight mistake (by a very small margin; see Figure~\ref{tab:DMDselectionNoEmbeddingScore}) on the second one, selecting \emph{Driveline} instead of \emph{Battery}. Since \emph{Driveline} has not changed version, it should have no influence. The fact that \emph{Driveline} is detected is likely a consequence of the insufficient quality of the reference \DMD{} model, which can be improved by adding an arbitrary variable, as illustrated in Figure~\ref{graf:dmdSelectionNoEmbeddingYYp}.

\begin{table}[H]
\centering
\begin{tabular}{rllrrrrl}
\toprule
    & Combinaison         & Status   &   RSS &   L1\_Aref &   L1\_Am &   Total Score & CPU   \\
\midrule
  3 & Motor . Driveline   & OK       &      846.720 &     8.652 &  18.740 &      8467.478 & 1.98s      \\
  0 & Battery . Motor     & OK       &      847.035 &    12.210 &  25.972 &      8470.730 & 1.99s      \\
  2 & Battery . Glider    & OK       &      850.546 &    11.276 &  34.000 &      8505.908 & 1.90s      \\
  1 & Battery . Driveline & OK       &      853.002 &     6.303 &  19.544 &      8530.280 & 1.98s      \\
  4 & Motor . Glider      & OK       &      853.384 &    14.653 &  36.040 &      8534.344 & 1.91s      \\
  5 & Driveline . Glider  & OK       &      855.590 &     8.103 &  23.224 &      8556.210 & 1.91s      \\
\bottomrule
\end{tabular}
\caption{Selection of the best combination of two modules, without embedding. The score is the value of the optimization objective; it combines the residuals and the absolute values of the matrix elements.}
\label{tab:DMDselectionNoEmbeddingScore}
\end{table}

\begin{table}[H]
\centering
\textbf{MixED-DMD identification: Motor . Driveline (11.67s)}

\vspace{1em}

\textbf{Reference (A)}\\
\resizebox{0.9\textwidth}{!}{
\begin{tabular}{lcccccccccccccc}
& \rotatebox{60}{B SOC} & \rotatebox{60}{B PowerLosses} & \rotatebox{60}{B EnergyLosses} & \rotatebox{60}{B VoltageAtTerminals} & \rotatebox{60}{M PowerOutput} & \rotatebox{60}{M PowerLosses} & \rotatebox{60}{M EnergyLosses} & \rotatebox{60}{M NetTorque} & \rotatebox{60}{D MotorSpeed} & \rotatebox{60}{D NetTractiveForce} & \rotatebox{60}{G PropellingEnergy} & \rotatebox{60}{G TractivePower} & \rotatebox{60}{G BrakingEnergy} & \rotatebox{60}{G Position} \\
      B SOC & \color{blue}{1.003} & -2.031 & - & 1.846 & -1.043 & -3.083 & - & -1.234 & 0.012 & -2.030 & - & -1.524 & -0.330 & - \\
      B PowerLosses & - & \color{blue}{1.042} & - & -0.046 & 0.032 & 0.024 & - & 0.013 & - & 0.021 & - & 0.026 & - & - \\
      B EnergyLosses & - & -0.334 & \color{blue}{1.000} & 0.225 & -0.070 & -0.437 & - & -0.135 & - & -0.276 & - & -0.132 & -0.063 & - \\
      B VoltageAtTerminals & - & 0.102 & - & \color{blue}{0.804} & 0.032 & -0.026 & - & - & - & 0.032 & - & 0.045 & -0.011 & - \\
      M PowerOutput & - & -0.086 & - & - & \color{blue}{0.856} & -0.085 & - & -0.086 & - & -0.107 & - & -0.075 & -0.010 & - \\
      M PowerLosses & - & - & - & -0.034 & 0.016 & \color{blue}{0.920} & - & 0.010 & - & -0.010 & - & 0.017 & - & - \\
      M EnergyLosses & - & -0.474 & - & 0.479 & -0.311 & -0.680 & \color{blue}{0.999} & -0.224 & - & -0.385 & - & -0.412 & -0.059 & - \\
      M NetTorque & - & 0.028 & - & -0.032 & 0.027 & 0.044 & - & \color{blue}{0.947} & - & 0.095 & - & 0.045 & 0.007 & - \\
      D MotorSpeed & - & -0.008 & - & -0.132 & -0.119 & -0.679 & - & -0.298 & \color{blue}{0.998} & -0.293 & - & -0.098 & -0.019 & - \\
      D NetTractiveForce & - & -0.027 & - & 0.031 & -0.036 & 0.006 & - & 0.020 & 0.008 & \color{blue}{0.894} & - & -0.053 & -0.009 & - \\
      G PropellingEnergy & - & -1.113 & - & 0.946 & -0.569 & -1.182 & - & -0.236 & 0.006 & -0.563 & \color{blue}{0.998} & -0.745 & -0.189 & - \\
      G TractivePower & - & 0.141 & - & -0.129 & 0.134 & 0.010 & - & 0.082 & - & 0.118 & - & \color{blue}{1.086} & - & - \\
      G BrakingEnergy & - & - & - & - & - & 0.022 & - & 0.016 & - & 0.011 & - & -0.005 & \color{blue}{1.078} & - \\
      G Position & - & -0.136 & - & 0.219 & -0.108 & -0.820 & - & -0.655 & - & -0.830 & - & -0.252 & -0.021 & \color{blue}{1.000} \\
      Variable impose vitesse cible & - & 0.053 & - & 0.105 & 0.141 & 0.634 & - & 0.236 & - & 0.241 & - & 0.110 & 0.030 & - \\
\end{tabular}
}

\vspace{1.2em}

\textbf{Motor}\\
\resizebox{0.9\textwidth}{!}{
\begin{tabular}{lcccccccccccccc}
& \rotatebox{60}{B SOC} & \rotatebox{60}{B PowerLosses} & \rotatebox{60}{B EnergyLosses} & \rotatebox{60}{B VoltageAtTerminals} & \rotatebox{60}{M PowerOutput} & \rotatebox{60}{M PowerLosses} & \rotatebox{60}{M EnergyLosses} & \rotatebox{60}{M NetTorque} & \rotatebox{60}{D MotorSpeed} & \rotatebox{60}{D NetTractiveForce} & \rotatebox{60}{G PropellingEnergy} & \rotatebox{60}{G TractivePower} & \rotatebox{60}{G BrakingEnergy} & \rotatebox{60}{G Position} \\
      B SOC & \color{blue}{-} & 2.817 & - & -0.846 & -0.213 & 2.575 & - & 0.159 & - & 0.304 & - & -0.366 & 0.969 & - \\
      B PowerLosses & - & \color{blue}{-0.235} & - & 0.122 & -0.092 & -0.081 & - & -0.036 & - & -0.015 & - & -0.077 & -0.005 & - \\
      B EnergyLosses & - & 0.373 & \color{blue}{-} & 0.018 & -0.217 & 0.338 & - & -0.030 & - & -0.011 & - & -0.267 & 0.162 & - \\
      B VoltageAtTerminals & - & -0.084 & - & \color{blue}{0.087} & 0.090 & -0.024 & - & 0.061 & - & 0.075 & - & 0.087 & -0.012 & - \\
      M PowerOutput & - & 0.446 & - & -0.259 & \color{blue}{0.387} & 0.147 & - & 0.186 & - & 0.141 & - & 0.343 & 0.015 & - \\
      M PowerLosses & - & - & - & 0.017 & - & \color{blue}{0.074} & - & - & - & -0.019 & - & -0.009 & -0.011 & - \\
      M EnergyLosses & - & 0.767 & - & -0.357 & 0.136 & 0.563 & \color{blue}{-} & -0.005 & - & 0.066 & - & 0.131 & 0.196 & - \\
      M NetTorque & - & -0.062 & - & 0.043 & -0.031 & -0.069 & - & \color{blue}{0.031} & - & 0.053 & - & -0.021 & -0.023 & - \\
      D MotorSpeed & - & -0.848 & -0.006 & 0.552 & -0.229 & 0.513 & -0.005 & 0.128 & \color{blue}{-} & 0.134 & - & -0.180 & -0.032 & - \\
      D NetTractiveForce & - & 0.070 & - & -0.052 & 0.049 & 0.038 & - & 0.008 & - & \color{blue}{-} & - & 0.043 & 0.031 & - \\
      G PropellingEnergy & - & 1.774 & - & -0.737 & 0.248 & 0.936 & - & -0.105 & - & 0.015 & \color{blue}{-} & 0.161 & 0.643 & - \\
      G TractivePower & - & -0.409 & - & 0.275 & -0.260 & -0.105 & - & -0.123 & - & -0.081 & - & \color{blue}{-0.230} & -0.027 & - \\
      G BrakingEnergy & - & - & - & - & - & -0.020 & - & -0.014 & - & -0.009 & - & 0.007 & \color{blue}{-0.100} & - \\
      G Position & - & -0.063 & - & 0.219 & -0.380 & 0.775 & - & 0.304 & - & 0.243 & - & -0.395 & -0.030 & \color{blue}{-} \\
      Variable impose vitesse cible & - & 0.741 & 0.005 & -0.482 & 0.166 & -0.484 & 0.005 & -0.101 & - & -0.110 & - & 0.127 & -0.015 & - \\
\end{tabular}
}

\vspace{1.2em}

\textbf{Driveline}\\
\resizebox{0.9\textwidth}{!}{
\begin{tabular}{lcccccccccccccc}
& \rotatebox{60}{B SOC} & \rotatebox{60}{B PowerLosses} & \rotatebox{60}{B EnergyLosses} & \rotatebox{60}{B VoltageAtTerminals} & \rotatebox{60}{M PowerOutput} & \rotatebox{60}{M PowerLosses} & \rotatebox{60}{M EnergyLosses} & \rotatebox{60}{M NetTorque} & \rotatebox{60}{D MotorSpeed} & \rotatebox{60}{D NetTractiveForce} & \rotatebox{60}{G PropellingEnergy} & \rotatebox{60}{G TractivePower} & \rotatebox{60}{G BrakingEnergy} & \rotatebox{60}{G Position} \\
      B SOC & \color{blue}{-0.006} & -0.681 & 0.006 & -0.864 & 0.708 & -1.442 & 0.007 & -0.315 & 0.007 & 1.066 & 0.006 & 1.083 & 0.246 & 0.006 \\
      B PowerLosses & - & \color{blue}{-0.112} & - & 0.068 & -0.051 & -0.034 & - & -0.018 & - & -0.026 & - & -0.041 & - & - \\
      B EnergyLosses & - & -0.227 & \color{blue}{-} & -0.065 & 0.032 & -0.373 & - & -0.097 & - & 0.148 & - & 0.097 & 0.048 & - \\
      B VoltageAtTerminals & - & -0.300 & - & \color{blue}{0.344} & -0.206 & 0.013 & - & -0.079 & - & -0.118 & - & -0.199 & 0.011 & - \\
      M PowerOutput & - & -0.089 & - & 0.197 & \color{blue}{-0.088} & 0.088 & - & -0.060 & - & -0.060 & - & -0.120 & 0.007 & - \\
      M PowerLosses & - & -0.060 & - & 0.064 & -0.042 & \color{blue}{0.013} & - & -0.035 & - & -0.012 & - & -0.041 & - & - \\
      M EnergyLosses & - & - & - & -0.268 & 0.215 & -0.175 & \color{blue}{-} & -0.120 & - & 0.171 & - & 0.271 & 0.043 & - \\
      M NetTorque & - & 0.024 & - & -0.018 & 0.016 & 0.058 & - & \color{blue}{0.086} & - & -0.065 & - & - & - & - \\
      D MotorSpeed & - & -0.914 & -0.005 & 0.575 & -0.259 & 0.334 & - & 0.052 & \color{blue}{-} & 0.075 & - & -0.228 & 0.026 & - \\
      D NetTractiveForce & - & -0.023 & - & 0.011 & - & -0.050 & - & -0.054 & - & \color{blue}{0.069} & - & 0.013 & - & - \\
      G PropellingEnergy & - & -0.146 & - & -0.414 & 0.296 & -0.727 & - & -0.450 & - & 0.181 & \color{blue}{-} & 0.439 & 0.144 & - \\
      G TractivePower & - & -0.138 & - & 0.086 & -0.081 & -0.060 & - & -0.013 & - & -0.034 & - & \color{blue}{-0.049} & - & - \\
      G BrakingEnergy & - & - & - & - & - & -0.021 & - & -0.013 & - & -0.007 & - & 0.006 & \color{blue}{-0.044} & - \\
      G Position & - & -0.305 & - & -0.127 & 0.172 & -0.162 & - & 0.358 & - & 0.584 & - & 0.282 & 0.012 & \color{blue}{-} \\
      Variable impose vitesse cible & - & 0.886 & 0.005 & -0.563 & 0.254 & -0.307 & - & - & - & -0.041 & - & 0.229 & -0.034 & - \\
\end{tabular}
}

\vspace{1.2em}

\textbf{Driveline.Motor}\\
\resizebox{0.9\textwidth}{!}{
\begin{tabular}{lcccccccccccccc}
& \rotatebox{60}{B SOC} & \rotatebox{60}{B PowerLosses} & \rotatebox{60}{B EnergyLosses} & \rotatebox{60}{B VoltageAtTerminals} & \rotatebox{60}{M PowerOutput} & \rotatebox{60}{M PowerLosses} & \rotatebox{60}{M EnergyLosses} & \rotatebox{60}{M NetTorque} & \rotatebox{60}{D MotorSpeed} & \rotatebox{60}{D NetTractiveForce} & \rotatebox{60}{G PropellingEnergy} & \rotatebox{60}{G TractivePower} & \rotatebox{60}{G BrakingEnergy} & \rotatebox{60}{G Position} \\
      B SOC & \color{blue}{0.005} & - & - & 1.390 & -1.207 & 1.268 & - & -0.267 & -0.018 & -1.620 & -0.006 & -1.833 & -0.762 & - \\
      B PowerLosses & - & \color{blue}{0.330} & - & -0.187 & 0.159 & 0.084 & - & 0.062 & - & 0.034 & - & 0.130 & - & - \\
      B EnergyLosses & - & 0.229 & \color{blue}{-} & 0.012 & 0.029 & 0.406 & - & 0.039 & - & -0.187 & - & -0.069 & -0.127 & - \\
      B VoltageAtTerminals & - & 0.396 & - & \color{blue}{-0.336} & 0.185 & 0.070 & - & 0.042 & - & 0.009 & - & 0.140 & 0.012 & - \\
      M PowerOutput & - & -0.266 & - & 0.080 & \color{blue}{-0.188} & -0.086 & - & -0.123 & - & -0.022 & - & -0.125 & -0.009 & - \\
      M PowerLosses & - & 0.071 & - & -0.069 & 0.050 & \color{blue}{-0.034} & - & 0.057 & - & 0.062 & - & 0.047 & 0.012 & - \\
      M EnergyLosses & - & -0.289 & - & 0.516 & -0.462 & 0.133 & \color{blue}{-} & -0.017 & - & -0.378 & - & -0.612 & -0.156 & - \\
      M NetTorque & - & 0.024 & - & -0.013 & - & 0.010 & - & \color{blue}{-0.106} & - & -0.111 & - & 0.015 & 0.011 & - \\
      D MotorSpeed & -0.005 & 1.413 & 0.006 & -0.973 & 0.654 & -0.410 & 0.006 & 0.099 & \color{blue}{-} & 0.102 & - & 0.587 & 0.016 & - \\
      D NetTractiveForce & - & -0.034 & - & 0.027 & -0.028 & - & - & 0.035 & - & \color{blue}{0.045} & - & -0.035 & -0.021 & - \\
      G PropellingEnergy & - & -0.575 & - & 0.980 & -0.852 & 0.661 & - & 0.088 & -0.007 & -0.649 & \color{blue}{-} & -1.085 & -0.526 & - \\
      G TractivePower & - & 0.485 & - & -0.296 & 0.282 & 0.105 & - & 0.151 & - & 0.033 & - & \color{blue}{0.191} & 0.022 & - \\
      G BrakingEnergy & -0.006 & 0.091 & 0.007 & -0.061 & 0.063 & 0.012 & 0.008 & 0.035 & - & 0.017 & 0.005 & 0.055 & \color{blue}{0.030} & 0.006 \\
      G Position & - & 0.624 & - & -0.107 & 0.071 & 0.057 & - & -0.387 & - & -0.428 & - & -0.075 & 0.046 & \color{blue}{-} \\
      Variable impose vitesse cible & 0.005 & -1.306 & -0.006 & 0.900 & -0.591 & 0.394 & -0.006 & -0.125 & - & -0.116 & - & -0.537 & 0.024 & - \\
\end{tabular}
}

\caption{Transition matrices of the best two-module combination found.}
\label{fig:DMD2MatrixAlpha}
\end{table}
These results, consistent with the \DMDc{} analysis, indicate that the workflow dynamics are not strictly linear, even when represented by multiple transition matrices rather than a single one. The intrinsic nonlinearities of the system remain and limit detection performance.

\subsection{Module Detection Using \DMDdetect{} with Latent Embedding}
\label{dmdavecembedding}
\begin{figure}[H]
  \centering
  \includegraphics[
      width=0.85\textwidth,
      trim={0cm 55cm 0cm 2cm},
      clip
  ]{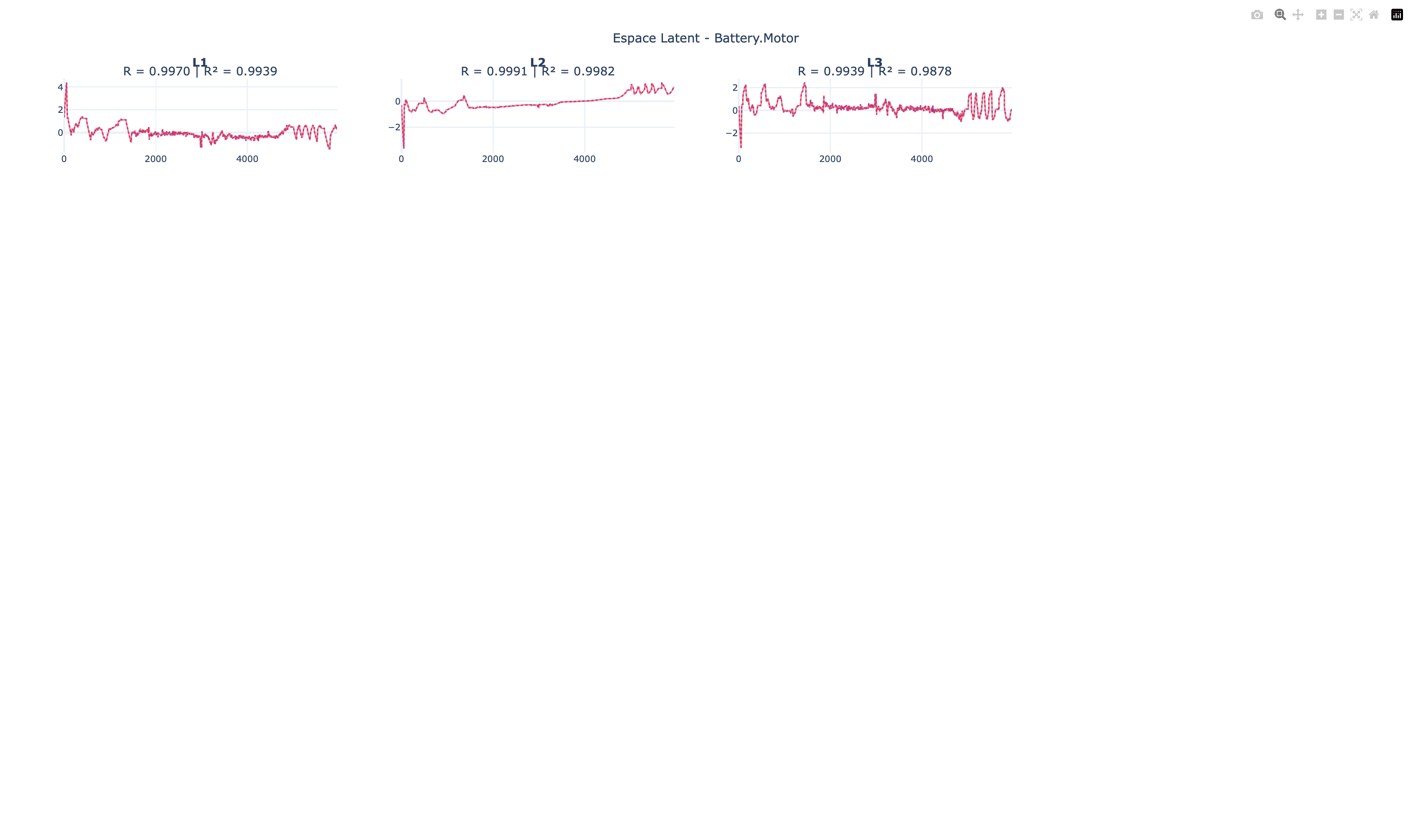}

  \vspace{0.6em}

  \includegraphics[
      width=0.85\textwidth,
      trim={0cm 1.8cm 0cm 2cm},
      clip
  ]{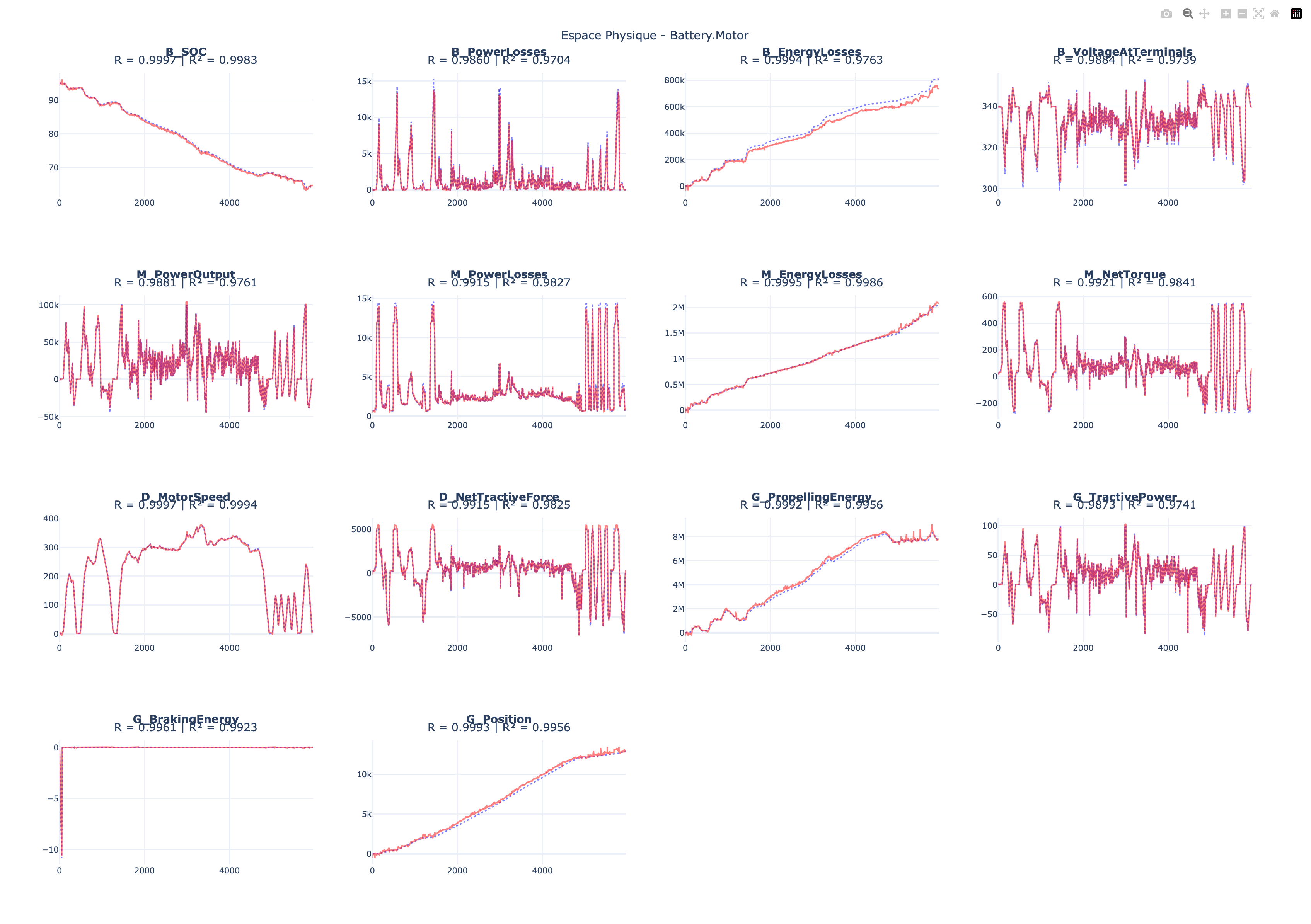}

  \caption{Embedding quality on the reference data table: three dimensions appear sufficient ($L_1$ to $L_3$). The models also appear satisfactory in the original space.}
  \label{dmdSelectionEmbeddingGraphiques}
\end{figure}

Embedding consists of applying non-linear transformations to the original variables in order to represent and simulate the complex vehicle behavior in a more compact and structured manner \cite{koopman1931hamiltonian,kutz2016dynamic}.
Embedding is systematically used in our method to minimize false detections caused by nonlinearities. \textbf{It is obtained using an \autoencoder{} trained on the reference data table}. The autoencoder was selected because it is simple to implement, easy to train, and independent of specific simulations such as $W_0(DoE)$. The dimension of the latent space is increased until a satisfactory model is obtained. The method correctly detects the two updated modules, as shown in Figures~\ref{dmdSelectionEmbeddingTableauCombinaisons}, \ref{dmdSelectionEmbeddingMatrices} and \ref{dmdSelectionEmbeddingGraphiques}. This amounts to assuming that the dynamic corrections represented by the matrices $A_{k,l}$ are linear in the autoencoder latent space. This assumption is consistent with our practical experience: module version changes only marginally modify the dynamics.

\begin{table}[H]
\centering
\begin{tabular}{lrrrrrl}
\hline
 Combinaison         &   $ \lambda_{Ref}$ &   $ \lambda_{Ref+T_1}$ &   $ \lambda_{Ref+T_2}$ &   $ \lambda_{Ref+T_1+T2+T1*T2}$ &    RSS & CPU   \\
\hline
 Battery . Motor     &     1.012 &    0.998 &    0.998 &             1.000 & 45.183 & 0.09s      \\
 Glider . Motor      &     1.010 &    1.000 &    0.997 &             1.001 & 45.495 & 0.08s      \\
 Driveline . Motor   &     1.003 &    1.008 &    0.996 &             1.000 & 45.529 & 0.08s      \\
 Battery . Driveline &     0.998 &    0.999 &    1.007 &             0.999 & 45.635 & 0.08s      \\
 Driveline . Glider  &     0.998 &    1.007 &    1.000 &             0.999 & 45.764 & 0.08s      \\
 Battery . Glider    &     1.000 &    0.997 &    0.999 &             1.000 & 45.808 & 0.08s      \\
\hline
\end{tabular}
\caption{Summary of two-module combinations. Model stability is not guaranteed for all observations, in particular those that use only the reference version of the workflow.}
\label{dmdSelectionEmbeddingTableauCombinaisons}
\end{table}

\begin{table}[ht!]
  \centering
  \scriptsize
  \textbf{Selected combination: Battery . Motor (0.49s)} \\[-0.3em]

  \begin{minipage}{0.42\textwidth}
    \centering \textbf{Reference (A)}
    \resizebox{0.9\textwidth}{!}{ 
    \begin{tabular}{lccc}
      & \rotatebox{60}{L1} & \rotatebox{60}{L2} & \rotatebox{60}{L3} \\
      L1 & \color{blue}{1.002} & -0.006 & - \\
      L2 & -0.013 & \color{blue}{1.007} & 0.023 \\
      L3 & - & - & \color{blue}{0.992} \\
    \end{tabular}}
  \end{minipage}\hfill
  \begin{minipage}{0.42\textwidth}
    \centering \textbf{Battery}
    \resizebox{0.9\textwidth}{!}{
    \begin{tabular}{lccc}
      & \rotatebox{60}{L1} & \rotatebox{60}{L2} & \rotatebox{60}{L3} \\
      L1 & \color{blue}{-0.009} & 0.010 & 0.014 \\
      L2 & 0.019 & \color{blue}{-0.011} & -0.038 \\
      L3 & - & - & \color{blue}{-} \\
    \end{tabular}}
  \end{minipage}

  \vspace{0.5em}

  \begin{minipage}{0.42\textwidth}
    \centering \textbf{Motor}
    \resizebox{0.9\textwidth}{!}{
    \begin{tabular}{lccc}
      & \rotatebox{60}{L1} & \rotatebox{60}{L2} & \rotatebox{60}{L3} \\
      L1 & \color{blue}{-0.015} & 0.018 & 0.016 \\
      L2 & 0.014 & \color{blue}{-0.010} & -0.022 \\
      L3 & -0.008 & - & \color{blue}{0.013} \\
    \end{tabular}}
  \end{minipage}\hfill
  \begin{minipage}{0.42\textwidth}
    \centering \textbf{Battery.Motor}
    \resizebox{0.9\textwidth}{!}{
    \begin{tabular}{lccc}
      & \rotatebox{60}{L1} & \rotatebox{60}{L2} & \rotatebox{60}{L3} \\
      L1 & \color{blue}{0.015} & -0.022 & -0.022 \\
      L2 & -0.026 & \color{blue}{0.013} & 0.044 \\
      L3 & 0.006 & - & \color{blue}{-} \\
    \end{tabular}}
  \end{minipage}

  \caption{Matrices $A$, $A_{Battery}$, $A_{Motor}$ and $A_{Battery.Motor}$ for the selected combination in the latent space. These matrices must be summed to obtain the different \DMD{} models used during the simulation.}
  \label{dmdSelectionEmbeddingMatrices}
\end{table}

\begin{figure}[H]
  \centering
  \includegraphics[
      width=0.90\textwidth,
      trim={0cm 1.8cm 0cm 4cm},
      clip
  ]{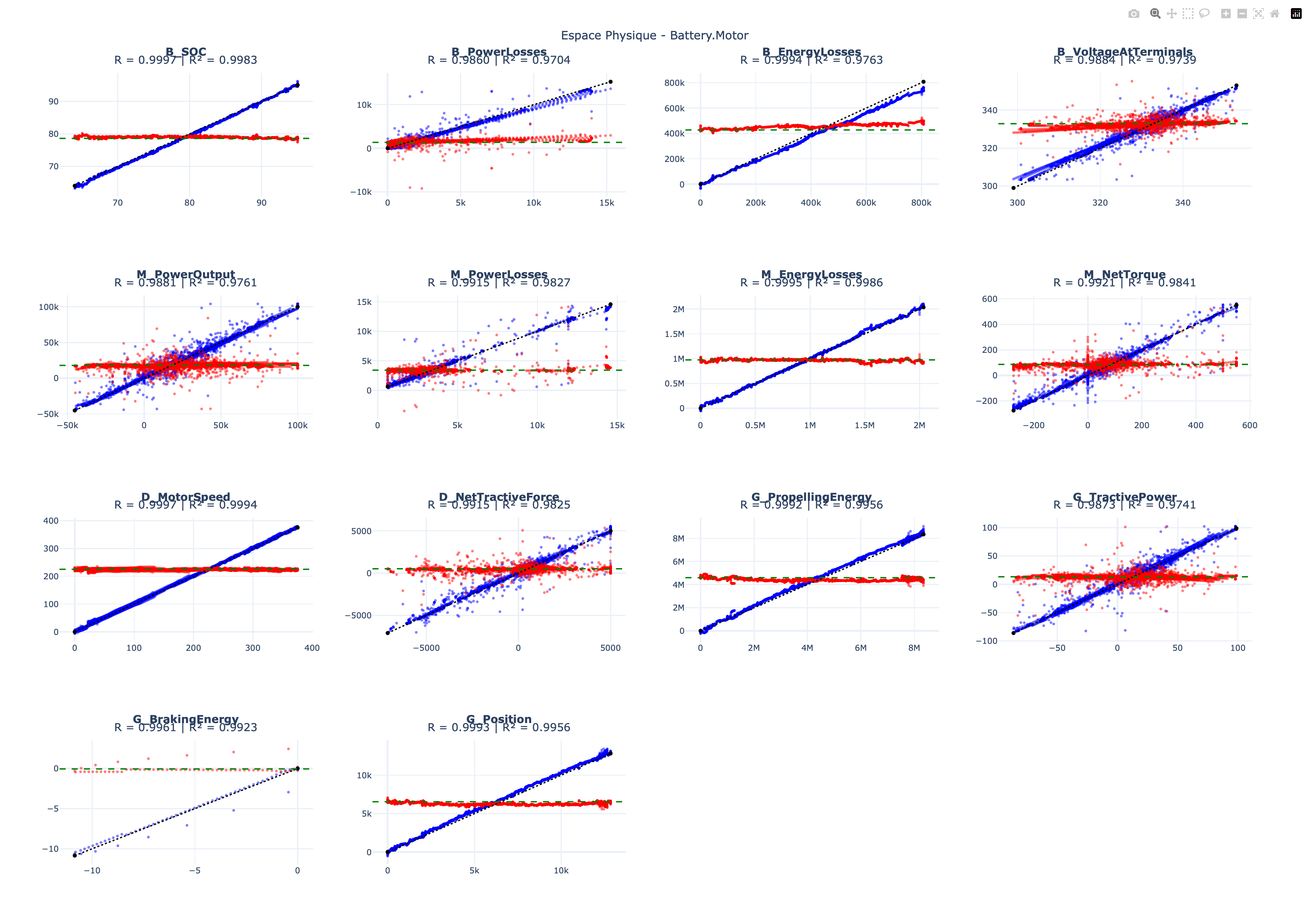}
  \caption{It can be observed that some observations exhibit large errors, which are not obvious in the time-series representations.}
  \label{dmdSelectionEmbeddingGraphiques2}
\end{figure}


\section{Two-Simulation Method with Reference (\NODYN{})} 
Modeling the dynamics in order to subsequently detect \enquote{faulty} modules is the most intuitive approach, but it is also the most challenging. Indeed, vehicle model dynamics are complex, and it is necessary each time to identify an appropriate embedding. There is no guarantee that the optimal embedding will systematically allow for linear modeling. Moreover, the workflow evolves throughout vehicle development projects, and studies become recurrent. Therefore, a detection method that is as automatic and as \enquote{frugal} as possible is required.

If an additional cost is acceptable—namely one specific additional simulation—it becomes possible to almost completely eliminate the need for dynamic modeling. This alternative method, although more costly, is used when there is doubt regarding the results of the first method.

\begin{figure}[H]
\centering

\begin{tikzpicture}[
    arrow/.style={->, thick}
]

\matrix (left) [matrix of nodes,
                row sep=6pt,
                column sep=10pt,
                draw] {
    $t$   & $v_1$ & $v_2$ & $v_3$ \\
    $t_1$ & $x_{11}$ & $x_{12}$ & $x_{13}$ \\
    $t_2$ & $x_{21}$ & $x_{22}$ & $x_{23}$ \\
    $t_3$ & $x_{31}$ & $x_{32}$ & $x_{33}$ \\
};

\node at ($(left.north)+(0,0.7)$) {$T_0(\text{DoE})$};

\matrix (right) [matrix of nodes,
                 row sep=6pt,
                 column sep=10pt,
                 draw,
                 right=5.5cm of left] {
    $t$   & $v_1$ & $v_2$ & $v_3$ \\
    $t_1$ & $\hat{x}_{11}$ & $\hat{x}_{12}$ & $\hat{x}_{13}$ \\
    $t_2$ & $\hat{x}_{21}$ & $\hat{x}_{22}$ & $\hat{x}_{23}$ \\
    $t_3$ & $\hat{x}_{31}$ & $\hat{x}_{32}$ & $\hat{x}_{33}$ \\
};

\node at ($(right.north)+(0,0.7)$) {$T_0^*$};

\draw[arrow]
  ($(left-3-4.center)!0.20!(right-3-2.center)$) --
  node[midway, above, text width=3.5cm, align=center]
       {row-by-row computation\\with $W_0$}
  ($(left-3-4.center)!0.80!(right-3-2.center)$);

\end{tikzpicture}

\caption{Row-by-row computation between $T_0(\text{DoE})$ and $T_0^*$. The simulation with $W_0(DoE)$ is first performed to obtain the data table $T_0(DoE)$. Each row of $T_0(DoE)$ is then used as the initial condition in a one time step simulation with $W_0$ to produce the corresponding row of $T_0^*$.}

\end{figure}
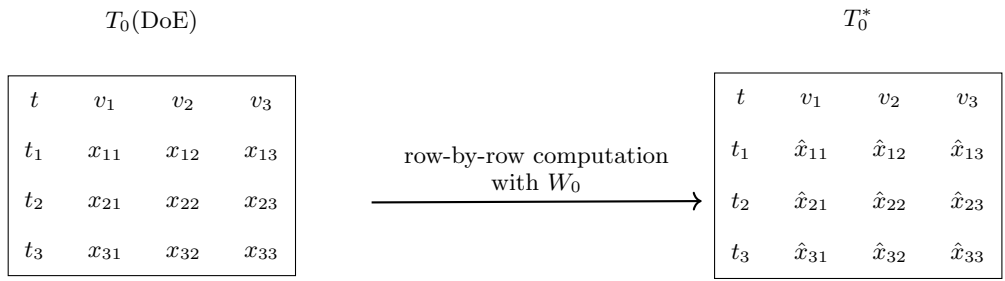

Detection is performed using linear regression:
\begin{enumerate}
    \item The table $T_0(DoE) - T_0^*$ is used as the response variable;
    
    \item The imposed variables, such as setpoint speed for example, are used as explanatory variables since they are identical in both simulations. Their variation between successive time steps may also be included;
    
    \item The experimental design over the modules is also used as explanatory variables (Boolean variables). For these Boolean variables, the model is necessarily linear with interactions (if the design allows them). If the experimental design is a full two-level factorial design, the model may include all interactions up to the highest order (equal to the number of modules). A significant interaction between multiple modules means that several modules must be updated simultaneously to explain the change in dynamics;
    
    \item The model may also include interactions between imposed variables and Boolean variables. However, the quality of $DoE$ depends on the imposed driving cycle and on the order in which the Boolean design is applied;
    
    \item The responses at the previous time step are also included as explanatory variables in order to represent the \enquote{remaining} dynamics. Indeed, the rows of the experimental design are repeated; for a given module combination, only the imposed variables vary at the input to explain the output responses. Since the imposed variables alone cannot fully explain the responses, at least the previous-step responses (and possibly interactions involving these responses) must be included.
\end{enumerate}

The detection capability of \NODYN{} is limited by points 4 and 5 above, as there is no guarantee that the added interaction terms are sufficient.

The following results correspond to the same test case as in the \DMDdetect{} method: the experimental design is a full factorial design with 16 combinations, repeated over time. Only the Battery and Motor modules were actually updated.

\begin{table}[ht]
\centering
\begin{tabular}{lrrrr}
\hline
                                &   Driveline &   Motor &   Battery &   Glider \\
\hline
 B\_EnergyLosses &       0.197 &   0.357 &     0.661 &    0.089 \\
 B\_PowerLosses  &       3.680 &   7.116 &    11.839 &    1.121 \\
 B\_SOC                &      -0.000 &  -0.000 &     0.000 &    0.000 \\
 B\_VoltageAtTerminals  &      -0.005 &  -0.006 &    -0.051 &   -0.008 \\
 D\_MotorSpeed    &       0.000 &  -0.000 &    -0.000 &   -0.000 \\
 D\_NetTractiveForce  &      -0.405 &  -0.308 &    -0.486 &    0.345 \\
 G\_BrakingEnergy        &      -0.001 &   0.000 &    -0.000 &   -0.000 \\
 G\_Position            &      -0.000 &   0.000 &    -0.000 &   -0.000 \\
 G\_PropellingEnergy     &       0.120 &  -0.502 &    -0.152 &   -0.196 \\
 G\_TractivePower       &      -0.004 &  -0.004 &     0.001 &    0.002 \\
 M\_EnergyLosses     &       0.025 &   1.203 &    -0.497 &    0.029 \\
 M\_NetTorque       &      -0.013 &   0.277 &    -0.284 &    0.013 \\
 M\_PowerLosses      &       0.078 &  22.098 &    -9.271 &   -1.147 \\
 M\_PowerOutput      &       7.013 &  26.833 &   -11.584 &   -3.419 \\
\hline
\end{tabular}

\caption{Module coefficients in the prediction of the responses. Only module coefficients are retained for final interpretation, as the analysis focuses exclusively on the modules.}

\label{tab:suppressDynamicsCoeff}
\end{table}

\begin{table}[H]
\centering

\begin{tabular}{lrrrr}
\hline
                                &   Driveline &   Motor &   Battery &   Glider \\
\hline
 B\_EnergyLosses &       0.334 &   0.064 &     0.027 &    0.540 \\
 B\_PowerLosses  &       0.347 &   0.059 &     0.039 &    0.692 \\
 B\_SOC                 &       0.393 &   0.007 &     0.704 &    0.796 \\
 B\_VoltageAtTerminals  &       0.532 &   0.437 &     0.000 &    0.203 \\
 D\_MotorSpeed    &       0.361 &   0.576 &     0.157 &    0.472 \\
 D\_NetTractiveForce  &       0.171 &   0.397 &     0.230 &    0.169 \\
 G\_BrakingEnergy        &       0.288 &   0.229 &     0.235 &    0.230 \\
 G\_Position             &       0.019 &   0.136 &     0.007 &    0.246 \\
 G\_PropellingEnergy     &       0.594 &   0.014 &     0.721 &    0.376 \\
 G\_TractivePower       &       0.246 &   0.245 &     0.827 &    0.404 \\
 M\_EnergyLosses     &       0.895 &   0.000 &     0.110 &    0.881 \\
 M\_NetTorque       &       0.878 &   0.021 &     0.029 &    0.865 \\
 M\_PowerLosses      &       0.983 &   0.000 &     0.111 &    0.741 \\
 M\_PowerOutput      &       0.490 &   0.041 &     0.459 &    0.687 \\
\hline
\end{tabular}

\caption{The coefficients are associated with an analysis-of-variance test, yielding a $p$-value. The smaller the $p$-value, the more influential the module is on the response.}

\label{tab:suppressDynamicsPval}
\end{table}

\begin{table}[H]
\centering

\begin{tabular}{lllll}
\hline
                                & Driveline   & Motor   & Battery   & Glider   \\
\hline
 B\_EnergyLosses & -           & 0.357   & 0.661     & -        \\
 B\_PowerLosses & -           & 7.116   & 11.839    & -        \\
 B\_SOC                 & -           & -       & -         & -        \\
 B\_VoltageAtTerminals  & -           & -       & -0.051    & -        \\
 D\_MotorSpeed    & -           & -       & -         & -        \\
 D\_NetTractiveForce  & -           & -       & -         & -        \\
 G\_BrakingEnergy        & -           & -       & -         & -        \\
 G\_Position             & -           & -       & -         & -        \\
 G\_PropellingEnergy     & -           & -0.502  & -         & -        \\
 G\_TractivePower       & -           & -       & -         & -        \\
 M\_EnergyLosses     & -           & 1.203   & -         & -        \\
 M\_NetTorque       & -           & 0.277   & -0.284    & -        \\
 M\_PowerLosses      & -           & 22.098  & -         & -        \\
 M\_PowerOutput      & -           & 26.833  & -         & -        \\
\hline
\end{tabular}
\caption{The analysis rule is very simple: all non-zero coefficients of modules with a $p$-value below $0.1$ are retained. Only the Motor and Battery modules are kept, which correspond exactly to the only modules that were actually updated.}

\label{tab:suppressDynamicsFinal}
\end{table}


The application to the test case is summarized in Figures~\ref{tab:suppressDynamicsCoeff}, \ref{tab:suppressDynamicsPval}, and \ref{tab:suppressDynamicsFinal}. The \enquote{Motor} and \enquote{Battery} modules are clearly the most influential and correspond exactly to the only modules that were actually modified in the simulation. 

In this test case, each response (more precisely, each deviation) is modeled using multiple linear regression (Python function \textbf{ols} from \textbf{statsmodels.regression.linear\_model}) based on the imposed variables, the module Boolean variables, and the responses at the previous time step, using the option \textbf{cov\_type='HC3'}.

\section{Conclusion}
We developed two detection methods, \DMDdetect{} and \NODYN{}, which address practical industrial cases efficiently. Existing quality tools make it possible to detect a difference in the outputs of a simulation workflow after model updates; our methods then make it possible to attribute this discrepancy to specific modules, at a minimal cost in terms of the number of simulations.

These non-intrusive methods, whose low cost enables systematic use, will become increasingly important with distributed simulations. Indeed, model creators/suppliers can now provide their models through network access (APIs) and update them silently.

The tools can also be applied to several simulations simultaneously, for example to validate multiple customer driving cycles at once, or to exploit several organizations of the design of experiments over the modules.

The presented results show that detection performance strongly depends on the quality of \DMD{}-based modeling. Moreover, a reduced predictive workflow model capable of extrapolation would be highly valuable to accelerate simulations, for instance for onboard use in the vehicle or for design through numerical optimization.
Improving the \DMD{} modeling of the reference data table, and in particular the embedding, therefore becomes our main objective. The current embedding is obtained using an autoencoder, which is very simple to implement. We are exploring several complementary directions:

\begin{itemize}
    \item Condensing all customer driving cycles into a single one. The objective is to \enquote{fill} the $(x_t,x_{t+1})$ space as uniformly as possible. This will increase the size of the \DMD{} models because variables will be less correlated, but it should reduce extrapolation errors;
    \item Representing control mechanisms more explicitly in the embedding, for example activation thresholds for air conditioning, trajectory control, \etc{}, using zero-order spline functions (see \cite{stone1997polymars,mounayer2025rraedy});
    \item Merging the two steps---embedding and modeling/detection---into a single stage as in \cite{rrae2024}, with the goal of forcing the embedding to enable linear modeling;
    \item Extending the reference embedding with additional embedding dimensions specific to each module.
\end{itemize}
\section*{Funding}
This work was supported by the CIFRE program of the Association Nationale de la Recherche et de la Technologie (ANRT) in collaboration with Renault Group.

\section*{Conflicts of Interest}
The authors declare that they have no conflicts of interest regarding the publication of this paper.

\section*{Data Availability Statement}
The data that support the findings of this study are available from the corresponding author upon reasonable request.

\section*{Author Contribution Statement}
Di Jiang and Yves Tourbier developed the methodology, performed the simulations, and wrote the original manuscript. 
Sebastian Rodriguez, Hervé Colin and Francisco Chinesta contributed to the conceptualization of the study, analysis of the results, and manuscript review and editing. 

\bibliographystyle{edpnum} 
\bibliography{biblio_all_parts} 

@book{zienkiewicz2005finite,
  title={The Finite Element Method: Its Basis and Fundamentals},
  author={Zienkiewicz, Olek C. and Taylor, Robert L.},
  year={2005},
  publisher={Elsevier Butterworth-Heinemann},
  edition={6}
}

@inproceedings{blochwitz2011fmi,
  title={The Functional Mock-up Interface for Tool Independent Exchange of Simulation Models},
  author={Blochwitz, Torsten and Otter, Martin and Arnold, Martin and Bausch, Christian and Clau{\ss}, Christoph and Elmqvist, Hilding and Junghanns, Andreas and Mauss, Jakob and Monteiro, Manuel and Neidhold, Thomas and others},
  booktitle={Proceedings of the 8th International Modelica Conference},
  year={2011}
}

@phdthesis{bordet2011modelisation0D,
  author       = {Nicolas Bordet},
  title        = {Modélisation 0D/1D de la combustion diesel : du mode conventionnel au mode homogène},
  type         = {Thèse de doctorat},
  school       = {Université d'Orléans},
  year         = {2011},
  language     = {french},
  note         = {⟨NNT : 2011ORLE2070⟩. ⟨tel-00717396⟩}
}

@phdthesis{hammadi2020reduction0D,
  author       = {Youssef Hammadi},
  title        = {Réduction d'un modèle 0D instationnaire et non-linéaire de thermique habitacle pour l’optimisation énergétique des véhicules automobiles},
  type         = {Thèse de doctorat},
  school       = {Université Paris Sciences et Lettres},
  year         = {2020},  
  language     = {french},
  note         = {Thermique [physics.class-ph]. ⟨NNT : 2020UPSLM027⟩. ⟨tel-03012674v2⟩},
}

@phdthesis{janiaud2011modelisation0D,
  author       = {Noëlle Janiaud},
  title        = {Modélisation du système de puissance du véhicule électrique en régime transitoire en vue de l’optimisation de l’autonomie, des performances et des coûts associés},
  type         = {Thèse de doctorat},
  school       = {Supélec},
  year         = {2011},
  language     = {french},
  note         = {⟨NNT : 2011SUPL0009⟩. ⟨tel-00660749v2⟩}
}

@online{siemens2023renault,
  author       = {{Siemens Digital Industries Software}},
  title        = {Renault Reduces Carbon Emissions and Accelerates Vehicle Development with Digital Twin},
  year         = {2023},
  url          = {https://resources.sw.siemens.com/en-US/case-study-renault-green},
  note         = {Consulté le 20 mai 2025}
}

@book{livreyto,
    author = "Daniel Benoist Sandrine Tourbier-Germain Yves Tourbier",
    title = "Plans d'expériences, construction et analyse",
    year = "1995",
    publisher = "TEC\&DOC Lavoisier"
}

@article{mandl1985use,
  title={Use of orthogonal Latin squares for testing compilers},
  author={Mandl, R.},
  journal={Communications of the ACM},
  volume={28},
  number={10},
  pages={1054--1058},
  year={1985},
  publisher={ACM}
}

@book{kuhn2013introduction,
  title={Introduction to Combinatorial Testing},
  author={Kuhn, D.R. and Kacker, R.N. and Lei, Y.},
  publisher={CRC Press},
  year={2013}
}

@online{mathworks2022bev,
  author    = {{MathWorks}},
  title     = {Simscape Battery Electric Vehicle Model},
  version   = {1.1.1},
  year      = {2022},
  url       = {https://github.com/mathworks/Simscape-Battery-Electric-Vehicle-Model/tree/v1.1.1},
  urldate   = {2025-06-02},
  note      = {GitHub repository}
}

@article{Proctor2016,
  title = {Dynamic Mode Decomposition with Control},
  volume = {15},
  ISSN = {1536-0040},
  url = {http://dx.doi.org/10.1137/15M1013857},
  DOI = {10.1137/15m1013857},
  number = {1},
  journal = {SIAM Journal on Applied Dynamical Systems},
  publisher = {Society for Industrial & Applied Mathematics (SIAM)},
  author = {Proctor,  Joshua L. and Brunton,  Steven L. and Kutz,  J. Nathan},
  year = {2016},
  month = jan,
  pages = {142–161}
}

@article{koopman1931hamiltonian,
  title={Hamiltonian systems and transformation in Hilbert space},
  author={Koopman, Bernard O.},
  journal={Proceedings of the National Academy of Sciences},
  volume={17},
  number={5},
  pages={315--318},
  year={1931},
  publisher={National Acad Sciences},
  note={Foundational article: introduces the idea that nonlinear dynamics can be represented as linear dynamics in the space of observables. Conceptual foundation of all modern extensions (EDMD, polynomial transformations, etc.).}
}

@article{kutz2016dynamic,
  title={Dynamic Mode Decomposition: Data-Driven Modeling of Complex Systems},
  author={Kutz, J. Nathan and Brunton, Steven L. and Brunton, Bing and Proctor, Jonathan},
  journal={SIAM Review},
  year={2016},
  note={Reference book and article on DMD and EDMD. Contains an entire chapter explaining why adding nonlinear observables (products, squares, interactions, splines) improves the approximation of Koopman dynamics.}
}

@article{rrae2024,
    title     = {Rank Reduction Autoencoders},
    author = {Mounayer, Jad and Rodriguez, Sebastian and Ghnatios, Chady and Farhat, Charbel and Chinesta, Francisco},
    journal   = {arXiv},
    year      = {2024},
    volume    = {abs/2405.13980},
    url       = {https://arxiv.org/abs/2405.13980},
    eprint    = {2405.13980},
    archivePrefix = {arXiv}
}

@article{stone1997polymars,
  author  = {Stone, Charles J. and Hansen, Mark H. and Kooperberg, Charles and Truong, Young K.},
  title   = {Polynomial Splines and Their Tensor Products in Extended Linear Modeling},
  journal = {The Annals of Statistics},
  year    = {1997},
  volume  = {25},
  number  = {4},
  pages   = {1371--1470},
  doi     = {10.1214/aos/1069362746},
  note    = {Introduction of POLYMARS. Method based on polynomial splines 
             with adaptive knot selection. Particularly effective for 
             high-dimensional problems and complex interactions.}
}

@article{Schmid2010,
  title={Dynamic mode decomposition of numerical and experimental data},
  author={Schmid, Peter J.},
  journal={Journal of Fluid Mechanics},
  volume={656},
  pages={5--28},
  year={2010},
  publisher={Cambridge University Press}
}

@book{Nemhauser1988,
  title={Integer and Combinatorial Optimization},
  author={Nemhauser, George L. and Wolsey, Laurence A.},
  year={1988},
  publisher={Wiley}
}

@book{Bertsimas1997,
  title={Introduction to Linear Optimization},
  author={Bertsimas, Dimitris and Tsitsiklis, John N.},
  year={1997},
  publisher={Athena Scientific}
}

@article{Diamond2016,
  title={CVXPY: A Python-Embedded Modeling Language for Convex Optimization},
  author={Diamond, Steven and Boyd, Stephen},
  journal={Journal of Machine Learning Research},
  volume={17},
  pages={1--5},
  year={2016}
}

@article{mounayer2025rraedy,
  title={RRAEDy: Adaptive Latent Linearization of Nonlinear Dynamical Systems},
  author={Mounayer, Jad and Rodriguez, Sebastian and Tomezyk, Jerome and Ghnatios, Chady and Chinesta, Francisco},
  journal={arXiv preprint arXiv:2512.07542},
  year={2025}
}


\end{document}